\documentstyle{article}

\newcommand{\Rr}{{\rm I\!R}}

\newcommand{\half}{\frac{1}{2}}

\newcommand{\C}{{\mathchoice {\setbox0=\hbox{$\displaystyle\rm C$}\hbox{\hbox
to0pt{\kern0.4\wd0\vrule  height0.9\ht0\hss}\box0}}
{\setbox0=\hbox{$\textstyle\rm C$}\hbox{\hbox
to0pt{\kern0.4\wd0\vrule  height0.9\ht0\hss}\box0}}
{\setbox0=\hbox{$\scriptstyle\rm C$}\hbox{\hbox
to0pt{\kern0.4\wd0\vrule  height0.9\ht0\hss}\box0}}
{\setbox0=\hbox{$\scriptscriptstyle\rm C$}\hbox{\hbox
to0pt{\kern0.4\wd0\vrule  height0.9\ht0\hss}\box0}}}} 

\newcommand{\Z}{{{\mathchoice  {\hbox{$\textstyle  Z\kern-0.4em  Z$}}
{\hbox{$\textstyle  Z\kern-0.4em  Z$}}
{\hbox{$\scriptstyle  Z\kern-0.3em  Z$}}
{\hbox{$\scriptscriptstyle  Z\kern-0.2em  Z$}}}}}

\newcommand{\Q}{{\mathchoice   {\setbox0=\hbox{$\displaystyle\rm
Q$}\hbox{\raise
0.15\ht0\hbox  to0pt{\kern0.4\wd0\vrule  height0.8\ht0\hss}\box0}}
{\setbox0=\hbox{$\textstyle\rm  Q$}\hbox{\raise
0.15\ht0\hbox  to0pt{\kern0.4\wd0\vrule  height0.8\ht0\hss}\box0}}
{\setbox0=\hbox{$\scriptstyle\rm  Q$}\hbox{\raise
0.15\ht0\hbox  to0pt{\kern0.4\wd0\vrule  height0.7\ht0\hss}\box0}}
{\setbox0=\hbox{$\scriptscriptstyle\rm  Q$}\hbox{\raise
0.15\ht0\hbox  to0pt{\kern0.4\wd0\vrule  height0.7\ht0\hss}\box0}}}}

\newcommand{\qed}{\ \ \rule{1ex}{1ex}}

\newtheorem{theo}{Theorem}[subsection]
\newtheorem{prop}[theo]{Proposition}
\newtheorem{defn}[theo]{Definition}

\newtheorem{cor}[theo]{Corollary}

\begin{document}

\begin{center}{\bf{\Large Level spacings for integrable quantum maps in genus zero 
 }}\footnote{
Partially supported by  NSF grant
\#DMS-9404637.}\\ 
\medskip

  Steve Zelditch\\
\medskip

{Johns Hopkins University, Baltimore, Maryland  21218}\\\medskip

March 1997

\end{center}

\addtolength{\baselineskip}{1pt} 

\setcounter{section}{-1}
\begin{abstract}

We study the pair correlation function   for a variety of 
completely integrable quantum maps in one degree of freedom. For simplicity we assume 
 that the classical phase space $M$ is the Riemann sphere $\C P^1 $  and that the classical map
is a fixed-time map $exp t \Xi_H$ of a Hamilton flow.  The quantization is then a unitary
N x N matrix $U_{t, N}$ and its pair correlation  measure $\rho^{(N)}_{2,t}$ gives the
distribution  of spacings between  eigenvalues  in an interval of length  comparable to the mean level spacing ($ \sim 1/N$).
 The physicists' conjecture (Berry-Tabor conjecture)
is that as $N \rightarrow \infty$, $\rho_{2,t}^{(N)}$ should converge to the pair correlation
function $\rho_{2}^{POISSON} = \delta_o + 1$ of a Poisson process. For any 2-parameter family
of Hamiltonians of the form $H_{\alpha, \beta} = \alpha \phi(\hat{I}) + \beta \hat{I}$ with
$\phi'' \not= 0$ we prove that this conjecture is correct for almost all $(\alpha, \beta)$
along the subsequence of Planck constants $N_m = [m (\log m)^4].$  In the addendum to this
paper [Z. Addendum], we further show that for polynomial phases $\phi$ the a.e. convergence
to Poisson holds along the full sequence of Planck constants for the Cesaro means of
$\rho_{2, t, \alpha, \beta}.$
\end{abstract}

\tableofcontents

\section{Introduction}

In this  paper we shall be  concerned with the fine structure of the spectra of some
completely integrable quantum maps in genus zero, that is, with quantizations  of
integrable   symplectic maps $\chi$ on the Riemann sphere  $M=
\C P^1$, equipped its standard
(Fubini-Study) form $\omega$  of integral area.  For any positive integer $N$, 
$(\C P^1, N \omega)$ is quantized by the Hilbert space ${\cal H}_N \cong \Gamma( L^N)$ of holomorphic sections
of the Nth power of the hyperplane section line bundle.
  The  quantum system then consists of a sequence  of unitary operators $\{U_{\chi, N}\}$ on 
 ${\cal H}_N$, a Hilbert space  of dimension N. For simplicity, we restrict attention to
quantizations $U_{t, N}$ of  Hamilton flows $\chi_t = \exp t \Xi_{H}$ where the Hamiltonian
 $H$ has no separatrix levels.   Our interest is  in the semiclassical asymptotics
 ($N \rightarrow \infty$) of the pair correlation
function $\rho_{2, t}^{(N)}$ and number variance $\Sigma_{2,t}^{(N)}(L)$ of the quantum systems. 
We first show that the time-averages  of these objects tend to the Poisson limits
$$\frac{1}{b-a} \int_a^b \rho_{2,t}^{(N)}dt \rightarrow \rho_2^{POISSON} = \delta_o + 1,\;\;\;\;\;\;\;\frac{1}{b-a}
\int_a^b\Sigma_{2,t}^{(N)}(L)dt \rightarrow L $$  as $N \rightarrow \infty$.  This is consistent
with the Berry-Tabor conjecture [B.T] that eigenvalues of completely integrable quantum systems behave
like random numbers (waiting times of a Poisson process).  However, it is only a weak test of
the conjecture since the averaging process itself induces a good deal of the randomness. A much
stronger test is whether the variance tends to zero.     For
special 2-parameter families of Hamiltonians $H_{\alpha, \beta} = \alpha \phi(\hat{I}) +
\beta \hat{I}$ (see \S 2 for the definition) we show that the variance
tends to zero at a power law rate. This implies that the individual systems are almost always
Poisson along a slightly sparse subsequence of Planck constants.  In an addendum [Z.Addendum]
we will further show that when $\phi$ is a polynomial, then the Cesaro means in $N$ of the pair
correlation function $\rho_{2,t, \alpha, \beta}^{(N)}$ tend almost always to $\rho_2^{POISSON}.$

Before describing the models and results more precisely, let us recall what the level spacings
problems are about.  In the quantization $H \rightarrow \hat{H}^{(N)}$ of Hamiltonians on
compact phase spaces $M$ of dimension $2 f$,
 the `Planck constant' is constrained to the values $h = 1/N$ and the 
spectrum  of $\hat{H}^{(N)}$ consists of $d_N \sim N^f$ eigenvalues
$\{\lambda_{N,j}\}$ in a bounded interval $[min H, max H]$.  Similarly, the spectrum
of a quantum map $U_{\chi, N}$ consists of $d_N$ eigenvalues $\{e^{i \theta_{N,j}}\}$
on the unit circle $S^1$.  The density of states in degree N
$$d \rho_1^{(N)} = \frac{1}{d_N} \sum_{j=1}^{d_N} \delta
(\lambda_{N,j}),\;\;\;\;\;\;\mbox{resp.}\;\;\;\;\;   d \rho_1^{(N)} = \frac{1}{d_N}
\sum_{j=1}^{d_N} \delta ( e^{i \theta_{N,j}})$$ has a well defined weak limit as $N
\rightarrow \infty$ which may be calculated by standard methods of microlocal
analysis (\S 2).  According to the physicists, there also exist asymptotic patterns in
the spectra  on the much smaller length scale  of the mean level spacing
$\frac{1}{d_N}$ between consecutive eigenvalues.  The pair correlation
function $\rho_2$, for instance, is the limit distribution of spacings between all
pairs of normalized eigenvalues $d_N \lambda_{N j}$.
 The
length scale $\frac{1}{d_N}$ is usually below the resolving power of micrlocal
methods. Hence the problem of rigorously determining the limit , or even of
determining whether it exists, has remained open for almost all quantum systems.  The
sole exceptions are the cases of almost all flat 2-tori [Sa.2] (see also [Bl.L]) and Zoll surfaces
[U.Z].  For other  rigorous results on level spacings for Laplacians on
 surfaces with
completely integrable geodesic flow, see [S][K.M.S][Bl.K.S].  

On the other hand, there exist numerous computer studies of eigenvalue spacings
in the physics literature which indicate that limit PCFs often
exist.  The following   conjectures give a rough guideline towards the
expected shape of the level spacings statistics: 
\medskip

$\bullet$ When the classical system is generic chaotic, $\rho_2 = \rho_2^{GOE}$ where
$\rho_2^{GOE}$ is the limit expected PCF for NxN random
matrices in the Gaussian orthogonal ensemble;

$\bullet$ When the classical system is generic completely integrable, $\rho_2 =
\rho_2^{POISSON}:= 1 +
\delta_0$.  That is,  at least on the level of the PCF, the
normalized spacings between eigenvalues behave like waiting times of a Poisson
process.  The term $\delta_0$ comes from the diagonal, while the term $1$ reflects
that any spacing between distinct pairs is as likely as any other. 

\medskip

These conjectures should not be taken too literally, and indeed cannot be since the
term `generic' is not precisely defined.  Our main purpose in this  article is
to test the Poisson conjecture against
 quantized Hamilton flows in one degree of freedom on the compact Kahler phase
space $\C P^1$.  Of course, they are necessarily completely integrable. It might also
be suspected that quantized Hamilton flows in one degree of freedom are necessarily trivial, but this is
not the case: as will be seen, PCF's of
 toral completely integrable systems on $\C P^1$  are
almost always Poisson along a slighty sparse subsequence of Planck constants.  
It should also be recalled that many of the model quantum
 chaotic systems, such as kicked tops and rotors and cat maps, take place in one degree of
freedom and 
 still defy rigorous analysis [Iz][Kea].

The quantum maps studied in this paper thus join a growing list of integrable
 quantum systems whose level spacings
have been shown rigorously to exhibit some degree of Poisson statistics. 
On the other hand, it is clear that not {\it all} of the quantum maps
in our 2-parameter families exhibit Poisson behaviour (see the Appendix
for a counterexample).   Rather, our results tend to corroborate
the (still rather vague) picture  that (some level of) Poisson statistics occurs  almost everywhere in 
an n-parameter family of non-degenerate integrable systems, but that a dense exceptional set of non-Poisson  systems occurs as well.  This probabilistic revision of the Berry-
Tabor conjecture seems to have been first proposed by Sinai 
in his study of a closely related lattice point problem [S]. It was also
stated clearly by Sarnak [Sa.2] in his proof that almost every flat 2-torus has a 
Poisson PCF but that a dense residual set of flat tori had no
well-defined PCF.  

 We should also emphasize that all of  the rigorous results
at the present time on
eigenvalue spacings of completely integrable systems only
pertain to the 2-level correlation function; neither the $k$- level
correlation functions for $k \geq 3$ nor  the  (nearest-neighbor) level spacings
distribution have been proved to be Poisson.    In this connection it is interesting
to recall a suggestion of  Sinai concerning
the `degree' of Poisson behaviour of typical members
 in an $n$-parameter family of quantum systems (e.g. as measured by the largest $k$ 
so that the $k$-level correlation
function is Poisson) : namely, that the typical degree could depend on the number 
$n$ of parameters in the system.   Thus our 2-parameter families 
(and the 2-parameter family of flat 2-tori) exhibit Poisson PCF's a.e., but it is unknown whether their higher-level
correlation functions or level spacings distribution are also Poisson.

Now let us be more precise about the models we will study.
In the usual Kahler quantization of  
$(\C P^1,  N \omega)$, 
 ${\cal H}_N$ may be identified with the space ${\cal P}_N$
of homogeneous holomorphic polynomials $f(z_1, z_2)$ of degree $N$ on $\C^2$.
  A classical Hamiltonian $H \in
C^{\infty}(\C P^1)$ is then quantized  as    
 a self-adjoint Toeplitz operator 
$$\hat{H}^{(N)} :=\Pi_N H \Pi_N: {\cal H}_N \rightarrow {\cal H}_N,\;\;\;\;\;\; 
\psi_{N,j} \rightarrow \Pi_N H\psi_{N,j}$$
where $\Pi_N$ is the (Cauchy-Szego) orthogonal  projection on ${\cal H}_N$  (\S 1).
Hence the quantum Hamiltonian system amounts to the eigenvalue problem:
$$\hat{H}^{(N)} \phi_{N,j} = \lambda_{N,j}\phi_{N,j}, \;\;\;\;\;-||H||_{\infty} \leq
\lambda_{N,1} \leq \lambda_{N,2}\leq \cdots \lambda_{N,N} \leq ||H||_{\infty}.$$

For a fixed value $h = \frac{1}{N}$ of the Planck constant, the distribution of
normalized spacings between all possible pairs of eigenvalues of $\hat{H}^{(N)}$ is
given by the Nth pair correlation `function' (measure)
$$d\rho_{2}^{(N)}(x) = \frac{1}{N}\sum_{i,j=1}^N \delta(x - N(\lambda_{N,i} -
\lambda_{N,j})).$$
Here, the eigenvalues are  rescaled, $\lambda_{N,j} \rightarrow N \lambda_{N,j}$
to have unit mean level spacing, i.e. so that    
$N (\lambda_{N,i+1} -  \lambda_{N,i} ) \sim 1$  on average  

 Our first result gives an explicit formula for the limit pair correlation function
$$d\rho_2 = lim_{N \rightarrow \infty} d\rho_{2}^{(N)}(x) $$
of a quantized Hamiltonian $\hat{H}^{(N)}$. It is of a similar nature to the pair
correlation function for a Zoll Laplacian ([U.Z]) and involves  dynamical
invariants of the  classical Hamiltonian flow $\exp t \Xi_H$  generated by $H$ on
the classical phase space.  Under some generic hypotheses (which will be stated
precisely in
\S 2), the formula is given by:
\medskip

\noindent{\bf Theorem A}~~~ {\it For the generic $H \in C^{\infty}(M)$,
the limit pair correlation function for the system $\hat{H}^{(N)}$ is given by:
$$\rho_{2}(f) = V \hat{f}(0) +  \sum_{k \in \Z} \sum_{\nu =1}^M
\sum_{j = 1}^{N(\nu)} \int_{(c_{\nu}, c_{\nu + 1})} \hat{f} (k T_j^{\nu}(E))
T_j^{\nu}(E)^2 d E$$
where:

\noindent(i) $V = vol\{(z_1, z_2) \in M \times M : H(z_1) = H(z_2)\};$\\
(ii) $\{ c_{\nu}\}$ is the set of critical values of $H$;\\
(iii) For a regular value $E \in (c_{\nu},c_{\nu+1})$, $H^{-1}(E)$ is a
 union of periodic orbits $\{\gamma_{\nu}^j\}_{j=1}^{N(\nu)}$ of $\exp t \Xi_H$  and 
$T_j^{\nu}(E)$ is the minimal positive period of the jth component.}
\medskip

It follows that the pair correlation function of quantized Hamiltonians in one degree  of
freedom is quite deterministic.  On the other hand, the eigenvalues of the associated
 quantized Hamiltonian flow are much more random.  

Before describing the results, let us recall the definition of a quantum map and of its
pair correlation function.  Suppose that $\chi_o$ is a symplectic map of a compact
symplectic manifold $(M, \omega).$  It is called {\it quantizable} if it can be lifted to a
contact transformation $\chi$ of the prequantum $S^1$ bundle $\pi: (X, \alpha)\rightarrow (M, \omega)$, 
where $d\alpha = \pi^* \omega$. We are mainly interested here in Hamiltonian flows $\chi_t$
and these are always quantizable (\S 1).  
We then define the quantization of the map $\chi_o$ to be
$$U_{\chi, N}: =\Pi_N\sigma_{\chi}
T_{\chi}\Pi_N : {\cal H}_N \rightarrow {\cal H}_N $$
where $T_{\chi}$ is the translation operator by $\chi$ on ${\cal H}_N$ and
 $\sigma_{\chi} \in C^{\infty}(M)$ is the `symbol', designed to make 
$U_{\chi, N}$ unitary.  All of the usual quantum maps, e.g. 
 `cat maps'  and kicker rotors  can be obtained
by this method [Z].

Since the eigenvalues lie on the unit circle, the rescaling to unit mean level
spacing leads to  the (period) pair correlation functions
$$d\rho^{(N)}_{2}(x) := \frac{1}{N}\sum_{i,j=1}^{N} \sum_{\ell \in \Z}
\delta (x - N(\theta_{N,i} - \theta_{N,j} - \ell N ))
= \frac{1}{N^2}\sum_{i,j=1}^{N} \sum_{\ell \in \Z}
\delta (\theta_{N,i} - \theta_{N,j} - 2\pi \ell - \frac{x}{N}).$$
The large $N$ behaviour of $d\rho^{(N)}_{2}$ is quite `random' in general
because the rescaling destroys the Lagrangean nature of $U_{\chi, N}.$
The question is whether there is some asymptotic pattern to the randomness.   

As mentioned above, we will restrict in this paper to a simple but reasonably representative case of the question, namely to 
 Hamiltonian flows generated by perfect Morse functions $H$ on
$ \C P^1$. The reason for restricting attention to $\C P^1$ is that it is the only symplectic
surface carrying a Hamiltonian $S^1$ action (i.e. it is a 'toric variety'), namely the usual
rotation of the sphere about an axis. The moment map is known as an action variable $I$. Any
perfect Morse function may be written as a function $H = \phi (I)$ of a global action variable.
Any  toral action can be quantized and in particular $I$ can be quantized as an operator
$\hat{I}^{(N)}$ whose spectrum lies on a one dimensional
 `lattice' $\{\frac{j}{N}: j = -N \dots N\}$.  It follows that $H$ is quantized
as an operator of the  form $\phi (\hat{I})$ and its flow can be quantized as a unitary group
of the form $U_{t, N} =\Pi_N e^{it {\cal N}
\hat{H}^{(N)}}\Pi_N$, where ${\cal N}$   equals to $N$ on ${\cal
H}_N$.   Hence the eigenangles have the form $t N \phi(\frac{j}{N})$ and the asymptotics of the PCF
can be reduced to the study  exponential sums of the form
$$S(N; \ell, t) = \sum_{j=1}^N e^{2 \pi i N \ell \phi(\frac{j}{N}) t}.$$
The Poisson conjecture is essentially that these exponential sums behave like random walks.
It is too difficult to analyse the individual exponential sums, but we can successfully
analyse some typical behaviour in families of such systems.  The first result is about the
mean behaviour as the $t$ parameter varies.
 \medskip

\noindent{\bf Theorem B (a)}~~~{\it Suppose $H: M \rightarrow \Rr$ is a perfect Morse
function on $\C P^1$.  Then:
the limit PCF $\rho_{2,t}$ and number variance
$\Sigma_{2,t}(L)$ for $U_{t, N}$   are  Poisson
on average in the sense:
$$\lim_{N \rightarrow \infty} \frac{1}{b-a} \int_a^b \rho_{2,t}^{(N)} dt =
\rho_2^{POISS}: =  1 +
\delta_0$$
and  
$$\lim_{N \rightarrow \infty} \frac{1}{b-a} \int_a^b \Sigma_{2,t}^{(N)}(L) dt =
\Sigma_2^{POISS}(L): = L$$
for any interval
$[a,b]$ of $\Rr.$}
\medskip

This result applies to the case of linear Hamiltonians and their Hamilton flows, whose pair
correlation functions are   clearly not individually Poisson (cf. \S 6).  For Poisson level
spacings, we  make some further assumptions on the Hamiltonian (or phase $\phi$).  Our
main result concerns the mean and variance of a 2-parameter family of Hamiltonians:

\medskip

\noindent{\bf Theorem B (b) }~~~{\it   Let $I$ denote an action variable on $\C P^1$ and 
let $H_{\alpha, \beta} = \alpha \phi(I) + \beta I$ with
$| \phi''| > 0$. Denote by 
 $\rho_{2;(t, \alpha, \beta)}^{(N)}$ the pair correlation measure for
the quantum map $U_{(t,\alpha, \beta), N} = exp(i t {\cal N} \hat{H}_{(\alpha,
\beta; N)} ).$  Then for any $t \not= 0$, any $T > 0$ and any $f \in {\cal S}(\Rr)$
with $\hat{f} \in C_o^{\infty}(\Rr)$ we
have $$\frac{1}{(2T)^2} \int_{-T}^T \int_{-T}^T
|\rho_{2;(t, \alpha, \beta)}^{N}(f) - \rho_{2}^{POISSON}(f)|^2 d\alpha
 d\beta = 0 (\frac{(\log N)^2}{N}).$$ }
\medskip

Thus, the mean pair correlation function in the family is Poisson and the variance tends to
zero at the rate $\frac{(\log N)^2}{N}.$  Following [Sa.2], we conclude:
\medskip

\noindent{\bf Corollary }~~~{\it Let $N_m =  [ m (\log m)^4]$.  Then, for almost
all $(\alpha, \beta)$, $\rho_{2;(t, \alpha, \beta)}^{N} \rightarrow \rho_2^{POISSON}$.}

It would be interesting to study quantizations of Hamilton flows in the case where the Hamilton
had saddle levels, as must happen if the genus is $> 0.$  It would also be interesting to
study completely integrable maps which are not Hamilton flows.  We hope to extend our methods
and results to these cases in the future.
\medskip

\noindent{\bf Acknowledgements}  This article was completed during visits to the Australian
National University and to the Newton Insitute.   In particular, we thank A.Hassell and Z.Rudnick
for comments on the proof of Theorem B and J.Marklof for discussions of incomplete theta series. 
We particularly thank M.Zworski for suggesting a simplification of the proof of Theorem 5.1.1

\section{Toeplitz quantization}

We now review the basics of Toeplitz quantization on $\C P^1$. 
 For futher background  on Kahler quantization, we refer to
[G.S]; for general Toeplitz quantization 
we refer to [B.G][Z].

Toeplitz quantization is a form of Kahler quantization, that is, of quantization
of symplectic manifolds in the presence of a holomorphic structure.  The basic
idea is that the quantum system is the restriction of the classical system to
holomorphic functions.  

To be more precise,  let $(M,\omega)$ be a compact Kahler manifold with integral
symplectic form.  Then there is a positive hermitian holomorphic line bundle $L\rightarrow M$ with
connection 1-form $\alpha$ whose curvature equals $\omega$.  In Kahler quantization,
the phase space $(M,\omega)$ is quantized as the sequence of finite dimensional
Hilbert spaces $\Gamma(L^{\otimes N})$ where $\Gamma$ denotes the holomorphic
sections.  In Toeplitz quantization, these spaces are put together as the Hardy
space $H^2(X)$ of CR functions on  the unit circle bundle $X$ in
$L^*$.

Thus, the setting for Toeplitz quantization is a compact contact manifold
$(X,\alpha)$ whose contact flow
\begin{equation} \phi^{\theta} : X \rightarrow X,\;\;\;\;\;\phi^{\theta *} \alpha = \alpha
\end{equation}
defines a free $S^1$-action with quotient a Kahler manifold $M$ whose Kahler form $\omega$
pulls back to $d \alpha.$   The
 Kahler structure on $M$ also induces a CR structure on $X$. The corresponding Hardy space
$H^2 (X)$ is the space of boundary values of holomorphic functions on the disc bundle of
$L^*$ which lie in $L^2(X)$.  The orthogonal 
(Cauchy-Szego) projector $\Pi : L^2(X) \rightarrow H^2(X)$  defines a
Toeplitz structure on $X$ in the sense of [B.G].

From the symplectic point of view, $H^2(X)$ is viewed as the quantization of the
symplectic cone
$$\Sigma = \{(x, r\alpha_x): r \in \Rr^+\} \subset T^*X-0.$$
To be precise, $\Pi$ is a Hermite Fourier integral operator with wave front set on the isotropic
submanifold $\Sigma^*:= \{(\sigma, -\sigma): \sigma \in \Sigma\} \subset T^*(X \times X).$  The CR structure corresponds
to a positive definite Lagrangean sub-bundle $\Lambda$ of $T\Sigma^{\bot}$, the symplectic normal bundle
of $\Sigma.$  The vector fields generating $\Lambda$ annihilate a ground state $e_{\Lambda}$ in
the quantization of the $T\Sigma^{\bot}$.  The symbol of $\Pi$ is the orthogonal projection
$\pi = e_{\Lambda} \otimes e_{\Lambda}^*$ onto this ground state.  For a detailed account
of these objects we refer to [B.G].

\subsection{Toeplitz quantization in genus zero}

In the case of $M=  \C P^1$, the contact manifold $X$ may be identified with $SU(2)$ and
the Hardy space $H^2(X)$ may be identified with the space of lowest weight vectors for the
right action of $SU(2)$ on $L^2(SU(2)).$  To make the Toeplitz theory more concrete, let
us recall how these identifications are made.
  
We first recall [G.H, \S I.3]  that the holomorphic line bundles over $\C P^1$ are all powers
$H^{\otimes N}$ of the hyperplane bundle $H \rightarrow \C P^1,$ whose fiber over
$V \in \C P^1$ is the space $V^*$ of linear functionals on the line thru $V$. The
Chern class of $H$ is the Fubini study form  $\omega_{FS}$, which generates
$H^2(\C P^1)$.  The holomorphic sections ${\cal H}_N$ of $H$ are given by the linear functionals $L$ on 
$\C^2$ by setting $s_L(V)= L|_{V}.$ More generally, the holomorphic sections of $H^{\otimes N}$
correspond to homogeneous holomorphic polynomials of degree $N$ on $\C ^2.$  

 The associated principal $S^1$ to $H$ is evidently the unit sphere $S^3 \subset \C^2$ which we 
 identify with $SU(2).$  As the boundary of the unit ball $B \subset \C^2$, it has a 
natural CR structure.  The associated Hardy space is the usual space of boundary values
of holomorphic functions on $B$.   Under the $S^1$ action  $e^{i \theta} (z_1,z_2) =
(e^{i \theta } z_1, e^{i \theta} z_2)$ of $S^3 \rightarrow \C P^1$, it is evident that the
holomorphic functions transforming by $e^{iN\theta}$ are given by homogeneous holomorphic
 polynomials of degree $N$. The Cauchy -Szego kernel is given by
 $\Pi_N(z, w) = \langle z, w\rangle ^N$.
 
We also recall that  the irreducible representations ${\cal H}_N$ of $SU(2)$ are given by its actions
on homogeneous polynomials.  By the Plancherel theorem,
 $L^2(SU(2)) = \oplus_{N=1}^{\infty} {\cal H}_N \otimes {\cal H}_N^*$.
The CR structure induced on $SU(2)$ by the identification $S^3 = \partial B \equiv SU(2)$ is
equivalent to that given by the lowering operator $L_{-}$ for the right action. The Szego projector $\Pi_N$ is then
the orthogonal projection onto ${\cal H}_N \otimes \psi_{N}$ where $\psi_N$ is the lowest
weight vector in ${\cal H}_N^*.$

Below we will often refer to an action operator $\hat{I}^{(N)}$ on $\C P^1.$  It may be identified
with the Planck constant $\frac{1}{N}$ times any generator (e.g. $L_z$) of a Cartan subgroup
of $SU(2)$. Thus its eigenvalues in ${\cal H}_N$ are the weights $\frac{j}{N}$.

\subsection{Quantum maps}

Symplectic maps $\chi_o$ on $\C P^1$ may be quantized 
by the Toeplitz method as long as $\chi$ lifts to a contact transformation $\chi$
of $(X,\alpha)$.    
The Toeplitz quantization is almost the translation operator $T_{\chi}$
by $\chi$ compressed to the Hardy space $H^2(X).$ Since $T_{\chi}$ does not usually
usually preserve $H^2(X)$, $\Pi T_{\chi} \Pi$  is not generally unitary; to unitarize
it one must equip it with a symbol.  In
[Z] it is described how to construct a symbol
$\sigma_{\chi}$ on $M$ for any quantizable symplectic map on any compact symplectic
$M$  so that  $$U_{\chi}:= \Pi \sigma_{\chi} T_{\chi}\Pi$$
 is unitary.  We will describe the symbol in some detail in \S 2.3. 
It automatically commutes with the $S^1$ action, so  is the direct
sum of the finite unitary operators, $U_{\chi,N}$ on $\Theta_N.$  We define
$U_{\chi, N}$ to be the quantization of $\chi_o$ with semiclassical parameter $1/N.$
Its
eigenvalues have the form
\begin{equation} Sp (U_{\chi,N}) = \{ e^{2 \pi i \theta_{N,j}}: j=1, \dots, d_N\}
\end{equation}
where $d_N = dim H^2_{\Sigma}(N)= N.$

Consider now the case of Hamilton flows $\chi_{o t} = \exp t \Xi_H$ on a general symplectic
manifold $(M,\omega).$ 

\begin{prop}  Hamilton flows are always quantizable. \end{prop}

\noindent{\bf Proof}  What needs to be proved is that $\exp t \Xi_H$ always
lifts to a  contact flow $\chi_t: X \rightarrow X$.  Equivalently that
$\Xi_H$ lifts to a contact vector field, say  $X_H.$  We prove this by lifting
$\exp t \Xi_H$ to a homogeneous Hamilton flow $\exp t \bar{\Xi}_H$ on the
symplectic cone
$\Sigma.$
Let us define the function
$$r : \Sigma \rightarrow \Rr^+,\;\;\;\;\;\;\; r(x, r \alpha_x) = r.$$
Thus, $\Sigma \cong X \times \Rr^+$, 
$X \cong \{r = 1\}$ and the $\Rr^+$ action is
generated by the vector ${\cal R} = r \frac{\partial}{\partial r}.$ 
 
The natural symplectic structure $\omega$ on $\Sigma$ is the restriction of
the canonical symplectic structure $\omega_{T^*X}$ on $T^*X$, which is homogeneous of
degree 1. Denoting by $\pi: X \rightarrow M$ the projection, we have:

\begin{equation} \omega =  r \pi^* \omega_M + dr \wedge \alpha.\end{equation}

The proof is simply that $\omega_{T^*X} = d \alpha_{T^*X}$ where $\alpha_{T^*X}$
is the action 1-form.  This equation restricts to $\Sigma$ where $\alpha_{T^*X} =
r \alpha$. Taking the exterior derivative gives the formula.

Now return to $H \in C^{\infty}(M)$ and consider the Hamiltonian
$\bar{H} (x, r) = r \pi^* H (x)$ on $\Sigma.$  It is homogeneous of degree 1
so its Hamilton vector field $\bar{\Xi}_{\bar{H}} = \omega^{-1}(d \bar{H})$ is
homogeneous of degree zero and then its Hamilton flow $\exp t \bar{\Xi}_{\bar{H}}$
is homogeneous of degree one. We claim that (i) the flow preserves $X$; and (ii)
its restriction $\chi_t$ to $X$ is a contact flow lifting $\exp t \Xi_H.$

 Indeed, we have
$$\iota_{ \bar{\Xi}_{\bar{H}}} \omega = d (r H) = r dH + H dr =
r \iota_{ \bar{\Xi}_{\bar{H}}} \omega_M +  \iota_{ \bar{\Xi}_{\bar{H}}} dr \wedge
\alpha$$
$$ = r \iota_{ \bar{\Xi}_{\bar{H}}} \omega_M - \alpha ( \bar{\Xi}_{\bar{H}}) dr
+ dr ( \bar{\Xi}_{\bar{H}}) \alpha.$$
Since all terms except $dr ( \bar{\Xi}_{\bar{H}}) \alpha$ are   $d\theta$-independent we
must have $ dr ( \bar{\Xi}_{\bar{H}}) =0.$  Here $d \theta$ denotes the vertical
one form of $X$. It is then obvious that
$$ - \alpha ( \bar{\Xi}_{\bar{H}}) = H,\;\;\;\;\;\;\;\;\iota_{ \bar{\Xi}_{\bar{H}}}
\omega_M = dH.$$
The second equation says that $\bar{\Xi}_{\bar{H}}$ projects to $\Xi_H$, i.e
$\bar{\Xi}_{\bar{H}}$ is a lift of $\Xi_H$.  Since
$${\cal L}_{\bar{\Xi}_{\bar{H}}} \alpha = \iota_{\bar{\Xi}_{\bar{H}}} d\alpha +
d (\iota_{\bar{\Xi}_{\bar{H}}} \alpha) = d H - d H $$
we also see that $\bar{\Xi}_{\bar{H}}$ is a contact vector field (here, ${\cal L}$
is the Lie derivative). \qed

\section{Density of States, pair correlation function and number variance}

\subsection{DOS}
 
Before considering the pair correlation function, we first describe the limit density of states (DOS)
 of quantum Hamiltonians and quantum maps in
the Toeplitz setting.  They can be easily determined from the trace formulae of [B.G]
and indeed the calculation is carried out in [Z, Theorem A]. Let us recall the results.

In the case of Hamiltonians, the DOS in degree N is defined by
\begin{equation} d\rho_1^{(N)} (\lambda) := \frac{1}{d_N} \sum_{j=1}^{d_N} \delta(\lambda -
\lambda_{N,j}).\end{equation}

By [B.G, Theorem 13.13] we have:

\begin{prop} The limit DOS
is given by 
$$\beta_o (f) = \int_M f(H) \omega\;\;\;\;\;\;(f \in C(\Rr)).$$ \end{prop}

In the case of quantum maps the DOS in degree N is defined by
\begin{equation} d\rho_1^{(N)} (z) := \frac{1}{d_N} \sum_{j=1}^{d_N} \delta(z - e^{2\pi
i\theta_{N,i}})\;\;\;\;\;\;\;\;\;\; z \in S^1. \end{equation}

The limit DOS $\beta_o$ is determined in [Z, Theorem A] and depends on whether the
classical map is periodic or aperiodic (i.e. the set of periodic points has measure zero).

\begin{prop} Let $\chi$ be a symplectic map of $(M, \omega).$

(a)  In the aperiodic case, $\beta = c_o d\theta$ where $c_o$ is the constant
$ (\int_M \sigma d\mu) $
with $\sigma$ the symbol of $U_{\chi}.$

(b) If $\chi^k = id$, then  $\beta$ is a linear combination of delta functions at the
kth roots of unity.  
\end{prop}

\subsection{The pair correlation function and number variance}

We recall here the definitions of the pair correlation function and number variance for
quantum maps $U_{\chi, N}$ in $f$ degrees of freedom.  Then $dim {\cal H}_N = d_N \sim N^f$ and
the spectrum has the form Sp$(U_{\chi, N}) = \{e^{i\theta_{Nj}}: j = 1, \dots, d_N\}.$  The
spectrum may be identified with the periodic sequence $\{\theta_{N j} + 2\pi n: n \in \Z,
j=1, \dots, d_N\}$ and then rescaled to given a periodic sequence of period N and mean
level spacing one: $\{ d_N \theta_{N j} + 2\pi n d_N: n \in \Z,
j=1, \dots, d_N\}$. 

\begin{defn} The pair correlation function of level N of a quantum map in $f$ degrees of
freedom     is the measure on $\Rr$ given by

\begin{equation} d\rho_2^{(N)}(x) := \frac{1}{d_N}\sum_{j,k=1}^{d_N} \sum_{n
\in
\Z}
 \delta (d_N (\theta_{N j} - \theta_{N j}) + 2\pi n d_N  - x)\end{equation}
The limit pair correlation function is then:
$$d\rho_{2}^{(N)} (x) = w-\lim_{N\rightarrow \infty} d\rho_{2}^{(N)} (x).$$ \end{defn}

  We often write the
integral $\int_{\Rr} f d\rho_2^{(N)} dx$ as 
$$\rho_2^{(N)}(f) =  \frac{1}{d_N}\sum_{j,k=1}^{d_N} \sum_{n
\in
\Z}
 f(d_N (\theta_{N j} - \theta_{N j}) + 2\pi n d_N ).$$
By the Poisson summation formula we have:
 \begin{equation} \rho_2^{(N)}(f) = \frac{1}{d_N^2}\sum_{\ell \in \Z} \hat{f}(
\frac{2
\pi \ell}{d_N}) \sum_{j,k = 1}^N 
 e^{i   \ell 
(\theta_{N,j} - \theta_{N,k})} = \frac{1}{d_N^2}\sum_{\ell } \hat{f}(
\frac{2 \pi \ell}{d_N})  | Tr U_{\chi, N}^{\ell}|^2. \end{equation}

A closely related spectral statistic is the number variance for
  the quantum
map $U_{\chi, N}$.  It is defined as follows  (cf. [Kea]):
 First, define the  density of the scaled eigenangles by 
$$ \rho_s^{(N)} (\theta) = \sum_{j = 1}^{d_N} \sum_{n \in \Z} \delta(\theta - d_N
\theta_{N j} + 2 \pi d_N n) =  \sum_{\ell \in \Z} [\frac{1}{d_N} \sum_{j=1}^{d_N} e^{2 \pi i
\ell \theta_{N,j}} e^{- 2\pi i \ell \theta/ d_N}] $$
$$ = 1 + \frac{2}{d_N} \sum_{\ell = 1}^{\infty} Tr U_{\chi, N}^{\ell} 
e^{- 2\pi i \ell \theta/ d_N}.$$

Then:

\begin{defn} The {\em number variance} of $U_{\chi, N}$ is defined by:

$$ \Sigma_{2}^{(N)} (L) = \frac{1}{N}\int_0^N |\int_{x - L/2}^{x + L/2}
\rho_s (y) dy - L|^2 dx $$
$$ = \frac{2}{\pi^2} \sum_{\ell = 1}^{\infty} \frac{1}{\ell^2}
\sin^2(\frac{\pi \ell L}{d_N}) |Tr U_{\chi, N}^{\ell}|^2 $$.\end{defn}

We observe that $\Sigma_2^{(N)} (L)$ is similar to $\rho_2^{(N)} (f)$ for $\hat{f} = \frac{sin x}{x}$
except that the  $\ell = 0$ term has been removed.

\subsection{Asymptotics of traces and exponential sums}

Before getting down to our specific models, let us make some general remarks about the
exponential sums $S(N, \ell) := Tr U_{\chi, N}^{\ell}$.  

First, the traces $Tr U_{\chi, N}^{\ell}$ have complete asymptotic expansions as $N \rightarrow
\infty$.  To state the results, we need some notation.  
Recall that the $S^1$ action on $X$ is denoted $\phi^{\theta}.$
For each $\ell$  put

\begin{equation} \Theta_{\chi, \ell} = \{ \theta_j  \rm {mod} 2\pi : \rm{Fix}(
 \phi^{\theta_j} \circ \chi) \not= \emptyset \}. \end{equation}

Assuming (as we will) that the maps have clean fixed point sets, the set
$\Theta_{\chi, \ell}$ is  finite and $ \rm{Fix}(
 \phi^{\theta_j} \circ \chi)$ is a conic submanifold of $\Sigma$. We denote its dimension by
$e_j$ and its base $ \rm{Fix}(
 \phi^{\theta_j} \circ \chi) \cap X$ by  $\rm{SFix}(
 \phi^{\theta_j} \circ \chi).$ 

The trace asymptotics then have the form:

\begin{prop} For each $\ell \not= 0$, 
$$Tr U_{\chi, N}^{\ell} \sim   \sum_{\theta_j 
\in \Theta_{\chi, \ell}} \sum_{r = 0}^{\infty} a_{\ell,j, r} N^{\frac{e_j - 1}{2} - r}
 e^{i N \theta_j}$$
for certain coefficients $ a_{\ell, j, r}$ . The leading coefficients are given by:
$$a_{\ell, j, 0} = \int_{SFix  \phi^{\theta_j} \circ \chi^{\ell}} d\mu_{\chi_{\ell}}$$
where $d\mu_{\chi_{\ell}}$ is the 
canonical density on $SFix ( \phi^{\theta_j} \circ \chi^{\ell})$.
\end{prop}

Before sketching the proof, let us describe in more detail the ingredients that go into
the principal coefficients.  Roughly speaking, the quantized map $U_{\chi, N}$ involves two pieces of
data in addition to the Szego projector: the map $\chi$ and the symbol $\sigma_{\chi}$.  We recall
from [Z] that the scalar principal symbol of $U_{\chi}$ is given by 
$\sigma_{\chi} = \langle \chi_* e_{\Lambda}, e_{\Lambda}\rangle^{-1}$
 where $e_{\Lambda}$
is the section of the 'bundle of ground states' corresponding to $\Pi.$  See [B.G, \S 11] for the
definition and properties of the ground states (normal Gaussians).
Since $\chi$ is rarely holomorphic, it will generally not commute with $\Pi$ and will
take $e_{\Lambda}$ to another ground state
$\chi_* e_{\Lambda}.$  After trivializing the 'bundle of ground states' the map $\chi_*$
may be described as follows: the derivative $d\chi$ defines a linear symplectic map
  $d \chi |_{\Sigma^{\bot}}$ on the symplectic normal bundle $T\Sigma^{\bot}$ of $\Sigma.$  The quantization
of the normal space is a space of Schwartz functions and the quantization of 
  $d \chi |_{\Sigma^{\bot}}$ is its image ${\cal M}( d \chi |_{\Sigma^{\bot}})$ under the 
metaplectic representation.  Then $\chi_* = {\cal M}( d \chi |_{\Sigma^{\bot}})$. 

As a Fourier integral Toeplitz operator,  $U_{\chi} \in I^o(X \times X, graph(\chi)')$, where
$$graph(\chi)' = \{ (\sigma,  - \chi(\sigma)): \sigma \in \Sigma\}.$$
Hence its symbol is a symplectic spinor on the graph.  Under the natural parametrization 
$j: \Sigma \rightarrow graph(\chi)$, the symbol may be viewed as a symplectic spinor on $\Sigma$ and 
 as discussed 
in [Z], unitarity of $U_{\chi}$ forces it to have the form
\begin{equation} j^*\sigma_{U_{\chi}} =  \langle \chi_* e_{\Lambda}, e_{\Lambda}\rangle^{-1} 
|d\sigma|^{\half} \otimes e_{\Lambda} \otimes e_{\Lambda}^* \end{equation}
where $|d\sigma|$ is the symplectic volume density on $\Sigma$.  Hence as a symplectic spinor
on $graph(\chi)$ it has the form
\begin{equation} \sigma_{U_{\chi}} =  \langle \chi_* e_{\Lambda}, e_{\Lambda}\rangle^{-1} 
( \chi_* |d\sigma|^{\half}\otimes  |d\sigma|^{\half}) \ \otimes  \chi_* e_{\Lambda} \otimes
  e_{\Lambda}^* \end{equation}
 
Now let us describe the symbolic aspects of the trace, assuming that $\chi$ has clean fixed point
sets.  Then each component of $Fix(\chi|_{\Sigma})$ carries a canonical density $dV$ and by inserting
the radial vector field ${\cal R}$ we get a canonical Liouville density $d\mu_{\chi} 
= i_{{\cal R}}dV$ on the base
$SFix(\chi)$ of the fixed point set.  Moreover, in the normal direction we have the symplectic
spinor factor $ \chi_* e_{\Lambda} \otimes  e_{\Lambda}^*.$  Taking the trace, we get the matrix
element $\langle \chi_* e_{\Lambda}, e_{\Lambda} \rangle$.  This cancels the scalar principal
symbol, so the symbolic trace just gives the Liouville volume of the fixed point set, as stated
above.  

Having discussed the ingredients in the above Proposition, we now sketch the proof.  For
further details we refer to 
 [B.G, Theorem 12.9].
\medskip

\noindent{\bf Sketch of Proof}    Consider the Fourier series
\begin{equation} \Upsilon_{\chi, \ell}(\theta) = \sum_{N=1}^{\infty} 
Tr U_{\chi, N}^{\ell} e^{i N \theta} = Tr U_{\chi}^{\ell} e^{i \theta {\cal N}}
\Pi \end{equation}
By the composition theorem of [B.G], $\Upsilon_{\ell}$ is a Lagrangean distribution
on $S^1$ with singularities at the values $\theta_j \in \Theta_{\chi, \ell}$ and with
singularity degrees beginning at $\frac{e_j}{2}.$  Hence:
\begin{equation} \Upsilon_{ \chi, \ell} \cong \sum_{\theta_j
\in \Theta_{\chi, \ell}}
\sum_{r = 0}^{\infty} a_{\theta_j, \ell, r} u_{ \frac{e_j}{2} - r}(\theta - \theta_j)
\end{equation}
where $u_m (\theta) = \sum_{N=1}^{\infty} N^{m- 1} e^{i N \theta}.$  Note that $u_m$ is a
periodic distribution with the same singularity at $\theta = 0$ as the homogeneous
distribution  $(\theta - \theta_j + i 0)^{\frac{e_j}{2} - r}$. 

The leading coefficients are given by the principal symbols of $\Upsilon_{\chi, \ell}(\theta)$
at the singularities.  From the symbol calculus of [B.G], the coefficients are given by the
symbolic traces described above.   The expansion stated in the Proposition
then follows by matching Fourier series. \qed
\medskip

\noindent{\bf Examples} Let us consider the form of the trace for quantum maps in one degree
of freedom:
\medskip

\noindent{\bf (a)}  Suppose that $\chi_t = \exp t \Xi_H$ is the fixed time map of
a Hamilton flow.  The fixed point set then consists of a finite number of
level sets $\{H = E_j(t)\}$, which must be periodic orbits of period $t$.
Pick a base point $m_j$ on each orbit and lift it to a point $x_j$ lying over
$m_j$ in $X$.  Then the lift of  $\{H = E_j(t)\}$ to $x_j$ is a curve which begins
and ends on $\pi^{-1}(m_j)$.  The difference in the initial and terminal
angle is of course given by the holonomy with respect to $\alpha.$  This
holonomy angle $\theta_j$ is independent of the choice of $m_j$ and of $x_j$ and  
$\Theta_{\chi, t}$ is the set of these holonomy angles. Then SFix $\chi_t
\circ \phi^{\theta_j}$ is two dimensional and hence $e = 3.$

As will be seen below, expressing $Tr U_{\chi_t, N}^{\ell}$ in terms of its eigenvalues
produces   a classical exponential sum in the completely integrable case.  The
 trace expansion produces a dual exponential sum involving dynamical data.  The principal
term can be obtained by applying the van der Corput
method [G.K][H.1], i.e.  Poisson summation followed by  stationary phase. However, the existence
of a complete asymptotic expansion would probably not be clear without the Toeplitz
machinery.  The trace expansion (or at least the principal term) is often used in  
 the physics literature under the name of the Gutzwiller trace formula.
Some remarks on its limitations are included below.  
\medskip

\noindent{\bf (b)} Suppose next that $\chi_o$ has only isolated non-degenerate fixed point sets.
Then $\chi$ fixes the entire fiber over each fixed point.  The only singularity
occurs at $\theta = 0$ and the dimension of Fix$\chi$ equals one.  Hence
$e = 2.$
\medskip

\noindent{\bf Special case: Quantum cat maps $U_{g,N}$}  These are the most familiar examples of quantum
maps, so let us see what the above proposition says about them.   In this case, the trace
can be calculated exactly and equals the character of the finite metaplectic representations.
For the exact calculation in Toeplitz setting, see [Z].  For the physics style calculation,
see [Kea].   Here we do the calculation asymptotically.

First we observe that if $g = \left( \begin{array}{ll} a & b \\ c & d \end{array} \right )$ is
hyperbolic then it has non-degenerate fixed points at $(x, \xi) \in \Rr^2/\Z^2$ such that
$g (x, \xi) \cong (x, \xi) mod \Z^2, $ i.e. at the points $(g - I)^{-1} \Z^2.$ The lifted
map $\chi_g$ actually has no fixed points.  However, $\chi_g \circ \phi^{\theta}$ has fixed points
if and only if $[(g\cdot (x, \xi), e^{2 \pi i (t + \theta)}) ] = [(x, \xi, \theta)]$, where
the bracket denotes the equivalence in the quotient space.  Since $g\cdot (x, \xi) =
(x, \xi) + (m,n)$ for some $(m,n) \in \Z^2$ we get that
  $[(x, \xi) + (m,n), e^{2 \pi i (t + \theta)}) ] = [(x, \xi, \theta)]$.  But
$[(x, \xi) + (m,n), e^{2 \pi i (t + \theta)}) ] = [(x, \xi), e^{2 \pi i (t + \theta )}
e^{i \pi m n} e^{i \pi \omega((x, \xi), (m,n))} ].$  It follows that 
$$\theta_{mn} = - \half ( m n + \omega((x, \xi), (m,n)).$$
Hence 
$$\Theta_{g, 1} = \{ - \half ( m n + \omega((x, \xi), (m,n)) : g (x, \xi) =
(x, \xi) + (m,n) \}.$$

The fixed point set of $\chi_g$ is therefore clean and has dimension one (the fibers over the fixed
points of $g$ on $\Rr^2/\Z^2$). The fixed points are non-degenerate in the directions normal
to the fibers, so the canonical density is given by $\frac{1}{\sqrt{det(I - g)}} d\theta$ where
$d\theta$ is the invariant measure on the fibers. Furthermore, the scalar principal symbol
of $U_{g, N}$ is given by $m(g) = \langle {\cal M}(g) e_{\Lambda}, e_{\Lambda} \rangle^{-1} =
2^{-\half} (a + d + i(b-c))^{\half}$ [Z, \S 5]. This factor is cancelled in the trace operation,
leaving
$$Tr U_{g, N} = \frac{1}{\sqrt{det(I - g)}} \sum_{(m,n) \in \Z^2 / (I - g) \Z^2} e^{i \pi (mn +
\omega((m,n), (1 - g) (m,n))}.$$

The traces $Tr U_{g,N}^{\ell}$ can be analysed in the same way  because in this example
 $U_{g,N}^{\ell} =
U_{g^{\ell}, N}.$  Indeed, $g \rightarrow U_{g,N}$ is the metaplectic representation of
$SL(2, \Z / N Z)$ realized on spaces of theta-functions [Z].  This connection makes it possible
to get  exact results on the level spacings of $U_{g,N}$ even though it is a quantum chaotic
map.  On the other hand, the results are quite different from the GOE behaviour conjectured
for generic quantum maps (cf. [Kea]). 
\medskip

\noindent{\bf Remarks on the trace expansion} We do not actually use the trace expansion
in this paper.  This is for two reasons.  The main one is that due to the eigenangle rescaling
the significant powers  $U_{\chi, N}^{\ell}$ of the quantum map in the formula for the PCF
 are those for which $\ell$ is
on the order of $N$.  But the trace expansion as it stands is only an asymptotic expansion as $N \rightarrow
\infty$ for fixed $\ell.$  

The`standard wisdom' regarding exponential sums (see [B.I][H.1][G.K]) is that the 
van der Corput method 
produces a dual exponential sum which is only simpler than the original (in general)
 when the dual  sum has  fewer terms. In the
case of the trace expansion, this
gain in simplicity only occurs if the number of fixed points of $\phi^{\ell}$  is
 slowly growing in $\ell$.  Only if the topological entropy of $\phi$ equals
zero can this be the case.  The most favorable case is probably that of completely integrable
systems, for which 
 the $\ell$th term has roughly $\ell$ critical
points.  Then the periodic orbit sum contains only  $\ell$ terms. But since the significant terms
occur when $\ell \sim N$ this is no simplification.

The trace expansion is not necessary in the integrable case because the eigenangles
can be written down explicitly by the WKB method.  This is the essential use of complete
integrability in this paper and also in [Sa.2].  The chaotic case is much more difficult because  
 no explicit formula for the eigenangles is known and because the trace formula leads to a dual sum
with an exponentially growing number of terms.

\section{PCF for Hamiltonians: Proof of Theorem A}

As mentioned in the introduction, the pair correlation problem for quantized
Hamiltonians is similar to that for Zoll surfaces as presented in [U.Z]. We can 
therefore follow the exposition in [U.Z, \S 3] to the extent possible and  omit details
which are essentially similar to the Zoll case.

There is no difference in this problem between the case of $M = \C P^1$ and the general
case of symplectic surfaces.  So in this section $M$ can be any closed symplectic surface.  However,
we will make some generic  simplifying assumptions
 on the Hamiltonian.  The first one is
\medskip

\noindent{\bf Assumption MORSE}: {\it $H :M \rightarrow \Rr$ is a Morse
function}.
\medskip

\noindent Let ${\cal E}$ denote the set of values of $H$ and let 
$c_1 < c_2 < \dots < c_{M+1}$ denote its set of critical values.   Then the
inverse image $H^{-1}(c_{\nu}, c_{\nu +1})$ consists of a finite number $N(\nu)$ of
connected components $X_j^{\nu}$ each diffeomorphic to $(c_{\nu}, c_{\nu +1}) \times
S^1.$ Hence for $E \in (c_{\nu}, c_{\nu +1})$, $H^{-1}(E) \cap X_j^{\nu}$ consists of
a periodic orbit $\gamma_j^{\nu}(E)$ of $exp t \Xi_H$.  Its minimal positive period
will be denoted by $T_j^{\nu}(E).$

For the sake of simplicity we will make a second assumption:
\medskip

\noindent{\bf Assumption Q}: {\it $T_j^{\nu}$ and $T_k^{\nu}$ are independent over
$\Q$ if $j \not= k.$}
\medskip

Under this (generic) condition, the manifold ${\cal P}$ of periodic points which arises
in the calculation of $\rho_2$ has the minimal number of components. 
  In the general case, the formula for $\rho_2$ is given by
a sum over connected components with dependent period functions [U.Z, Theorem 3.3]. 
In the simpler case (which exhibits all of the ideas) we will prove:
\medskip

\begin{theo} With assumptions MORSE and Q on $H$, the limit pair
correlation function is given by
$$\int_{\Rr} f(x) d\rho_2(x) = V \hat{f}(0) + \sum_{k \in \Z} \sum_{\nu =1}^M
\sum_{j = 1}^{N(\nu)} \int_{(c_{\nu}, c_{\nu + 1})} \hat{f} (k T_j^{\nu}(E))
T_j^{\nu}(E)^2 d E$$
for any $f$ such that $\hat{f} \in C^{\infty}_o(\Rr).$\end{theo}

\noindent{\bf Proof:} To determine the asymptotics of the sequence $\rho_2^{(N)}(f)$
we form the generating function
$$\Upsilon_f (\theta) := \sum_{N=1}^{\infty} \rho_2^{(N)}(f) e^{i N \theta}.$$
We wish to show that $\Upsilon_f$ is a classical Hardy-Lagrangean distribution on
$S^1$.  The asympotics of $\rho_2^{(N)}(f)$ can then be determined from the singularity
data of $\Upsilon_f$.
 
To show that $\Upsilon_f (\theta)$ is a Lagrangean distribution, we write it as
the trace of a Toeplitz type Fourier Integral operator,  defined
as follows: We form the product manifold $X \times X$ and consider
the product Szego projector $$\Pi \otimes \Pi : L^2 (X \times X) \rightarrow
H^2(X) \otimes H^2(X)$$
and the diagonal projector
$$\Pi_{diag} := \bigoplus_{N=1}^{\infty} \Pi_N \otimes \Pi_N :  L^2 (X \times X)
\rightarrow \bigoplus_{N=1}^{\infty} \Theta_N \otimes \Theta_N.$$
We next observe that
$$ \rho_2^{(N)}(f) = Tr \Pi_N \otimes \Pi_N  \int_{\Rr} \hat{f}(t) e^{i t N (\hat{H}_N
\otimes I - I \otimes \hat{H}_N)} dt.$$
Noting that $N$ is the eigenvalue of the number operator ${\cal N}$ we can rewrite
this in the form
$$ Tr \Pi_N \otimes \Pi_N  \int_{\Rr} \hat{f}(t) e^{i t  ( {\cal N} \hat{H}_N
\otimes I - I \otimes {\cal N} \hat{H}_N)} dt.$$ 
The generating function is then given by
$$\Upsilon_f(\theta) = \sum_{N=1}^{\infty} Tr e^{i\theta [{\cal N} \otimes I } 
\Pi_N \otimes \Pi_N  \int_{\Rr}
\hat{f}(t) e^{i t  ( {\cal N} \hat{H}_N \otimes I - I \otimes {\cal N} \hat{H}_N)}
dt =$$
$$=    Tr \Pi_{diag} e^{i\theta [{\cal N} \otimes I ]}  
\int_{\Rr} \hat{f}(t) e^{i t  ( {\cal N} \hat{H} \otimes I - I \otimes {\cal N}
\hat{H})} dt.$$
We recall here that $\hat{H} = \Pi H \Pi$ where $H$ is the pull back to $X$ of
the function so denoted on $M.$
We now have to analyse each operator which occurs under the trace sign.
\medskip

\noindent{\bf (a) $\Pi_{diag}$}: This operator is the composition of the product
Szego projector $\Pi \otimes \Pi$ with the full diagonal weight projection
$$P_{diag} : L^2(X \otimes X) \rightarrow L^2_N(X) \otimes L^2_N(X)$$
where $L^2_N(X)$ is the eigenspace of ${\cal N}$ of eigenvalue $N$.

 From 
[G.S.2][U.Z] it follows that $P_{diag}$ is a Fourier Integral operator in the class
$I^{0} (X^{(2)} \times X^{(2)}, \tilde{\Gamma})$ where $X^{(2)} = X \times X$
and where $\tilde{\Gamma}$ is the flow-out of the coisotropic cone 
$$ \tilde{\Theta} = \{( \zeta_1, \zeta_2) \in T^*(X^{(2)}) : H (\zeta_1) -
H(\zeta_2) = 0 \}.$$
That is, let $\Phi^t = \exp t \Xi_H \times \exp -t \Xi_H$ denote the Hamilton
flow generated on $T^*(X^{(2)})$ by $H (\zeta_1) -
H(\zeta_2).$ Then in a well-known way, the map
$$i_{\tilde{\Theta}} : \Rr \times \tilde{\Theta} \rightarrow T^*(X^{(2)}) \times
T^*(X^{(2)}),
\;\;\;\;\;\;\;\; (t, \zeta_1, \zeta_2) \rightarrow ( (\zeta_1, \zeta_2),
\Phi^t (\zeta_1, \zeta_2) )$$
defines a Lagrange immersion with image equal (by definition) to the flow out
$\tilde{\Gamma}$.

On the other hand $\Pi \otimes \Pi$ is the exterior tensor product of two
Toeplitz (hence Hermite type Fourier Integral) operators.  According to
[B.G, Theorem 9.3], we therefore have
$$\Pi \otimes \Pi = \alpha + \beta$$
with 
$$\alpha \in  I^o ( X^{(2)} \times X^{(2)}, \Sigma \times \Sigma)$$
and with $WF(\beta)$ contained in a small conic neighborhood ${\cal C}$ of
$\Sigma \times 0\cup 0 \times \Sigma.$  Moreover, the symbol of $\Pi \otimes
\Pi$ is given by 
$$\sigma(\Pi \otimes \Pi) = \sigma(\Pi) \otimes \sigma(\Pi)$$
on $\Sigma \times \Sigma - {\cal C}.$  Hence $\Pi \otimes \Pi$ is essentially
a Toeplitz structure on the symplectic cone $\Sigma \times \Sigma \subset
T^*(X^{(2)}).$  The complication due to ${\cal C}$ will ultimately prove to
be irrelevant in the analysis of the trace since it will not contribute to the
singularities along the diagonal. Hence we can (and will) pretend that this 
component of $WF(\Pi \otimes \Pi)$ does not occur.

By the composition theorem for Hermite and ordinary Fourier Integral operators
[B.G, Theorem 7.5] it follows that (modulo the term $P_{diag} \circ \beta$) 
$$\Pi_{diag} \in I^o(X^{(2)} \times X^{(2)}, \Gamma)$$
where $\Gamma$ is the flowout Lagrangean in $\Sigma \times \Sigma$ for the
co-isotropic subcone $\Theta := \tilde{\Theta} \cap \Sigma \subset \Sigma.$
That is, the map
$$i_{\Theta}:= i_{\tilde{\Theta}}|_{ \Rr \times \Theta}: \Rr \times \Theta 
\rightarrow \Sigma \times \Sigma$$ is a Lagrange immersion with respect to the
symplectic cone $\Sigma \times \Sigma$ and $\Gamma$ is its image; of course it
is only an isotropic immersion with respect to $T^*(X^{(2)} \times X^{(2)}).$
\medskip

\noindent{\bf (b) $e^{i\theta [{\cal N} \otimes I ]}$:} This
operator does not require a fancy analysis since ${\cal N}$ is simply the
differentiation operator by the generator $\frac{\partial}{\partial \theta}$
 of the contact flow.  Hence
$e^{i\theta [{\cal N} \otimes I]}$ is the translation operator
$F(x,y) \rightarrow F(\phi^{\theta} (x), y)$ by $\phi^{\theta} \times id$ on 
$X^{(2)}.$
\medskip

\noindent{\bf (c) $ \Pi \otimes \Pi \int_{\Rr} \hat{f}(t) e^{i t  ( {\cal N} \hat{H}
\otimes I - I
\otimes {\cal N} \hat{H})} dt$ }: Here we have inserted the factor $\Pi \otimes
\Pi$, as we may, to simplify the discussion.

Since $[{\cal N}, \Pi] = 0 = [{\cal N}, \hat{H}]$, we may remove the projection
$\Pi$ from the exponent.  Then 
the unitary under the integral is given by 
$$e^{it  H {\cal N} } \otimes e^{-it  H {\cal N} }.$$  
Each factor is the exponential of a pseudodifferential operator of real principal
type and is therefore Fourier Integral.  It follows again from the composition
theorem [B.G, Theorem 7.5] that  
$$\Pi e^{it  H {\cal N} } \in I^{-\frac{1}{4} } ( X \times X,  C \cap
\Rr \times \Sigma
\times \Sigma),\;\;\;\;\;C := \{ (t, \tau, x, \xi, y, \eta): \tau + \sigma_{{\cal
N}}(x, \xi) H ( x)  = 0, \psi^t (x,\xi) = (y,\eta) \}$$
where $\psi^t$ is the Hamilton flow on $T^*X$ generated by  $\sigma_{{\cal
N}}(x, \xi) H .$   Note that $\sigma_{{\cal N}(x,\xi)} =
 \langle \xi, \frac{\partial}{\partial \theta} \rangle$,
which generates the lift of the central circle action on $X$ to $T^*X$.  Also,
the Hamilton flow of $H$ on $T^*X$ is the two-fold lift of the Hamilton flow
of $H$ on $M$: first from $M$ to $X$  and then from $X$
to $T^*X.$ Since   the Hamilton vector field of $\sigma_{{\cal N}(x,\xi)} H$ on $T^*X$
is given by 
$$ \Xi_{\sigma_{{\cal N}(x,\xi)} H} = H \Xi_{\sigma_{{\cal N}(x,\xi)} } + 
\sigma_{{\cal N}(x,\xi)} \Xi_H$$
and the Lie bracket of the two terms is zero, we have
$$\psi^t = \exp t H \Xi_{\sigma_{{\cal N}(x,\xi)} } \circ \exp t \sigma_{{\cal
N}} (x,\xi) \Xi_H.$$

The isotropic cone $C \cap \Sigma \times \Sigma$ can be parametrized by 
$$i_{C_{\Sigma}} : \Rr \times \Sigma \rightarrow \Rr \times \Sigma \times
\Sigma,\;\;\;\;\;(t, \zeta) \rightarrow (t, \sigma_{{\cal N}} H (\zeta), 
\zeta, \psi^t(\zeta)).$$
The symbol of $ \Pi \otimes \Pi \int_{\Rr} \hat{f}(t) e^{i t  ( {\cal N} \hat{H}
\otimes I - I
\otimes {\cal N} \hat{H})} dt$ may then be identified with the spinor 
$\hat{f}(t) |dt|^{\half} \otimes \sigma_{\Pi}.$
\medskip

Putting the above together we see that up to the factor of $P_{diag}$ the operator
under the trace is a Hermite Fourier Integral operator associated to the graph of
$\phi^{\theta} \times id \circ \psi^t \times \psi^{-t}$ on $\Sigma \times
\Sigma.$ The effect of the $P_{diag}$ factor is to reduce this torus action to the
quotient of $\Sigma \times \Sigma$ by the diagonal contact flow. 

The remainder of the calculation is similar to the Zoll case, so we will just summarize the
key ideas and refer to [U.Z] for details.
First,  the singularities of $\Upsilon$ are caused by the periodic orbits of
the reduced flow $\exp t \Xi_H \times \exp - t \Xi_H$ 
 on the characteristic variety ${\cal V}: = \{ H(z_1) - H(z_2) = 0\} \subset M
 \times M.$  The singularities occur at the holonomy angles of the lifts of these periodic orbits to the
prequantum $S^1$ bundle of $M \times M.$  Since the periodic orbits of the product (or
difference) flow are
products of periodic orbits of the factors, the holonomy of the periodic orbits are ratios of
holonomies of the factors.  In particular, if for each period there is just one periodic orbit,
then the ratio is always one and there is only a single singularity at $\theta = 0.$  The
general case is discussed in detail in the analogous situation of [U.Z].

The set of relevant periodic points and their periods is given by
$${\cal P} = \{(z_1, z_2, T) \in {\cal V} \times \rm{supp} \hat{f} :  
exp T \Xi_H (z_j) = z_j \}.$$
Observe that each component of ${\cal P}$ is of dimension 3: the $T = 0$
component equals ${\cal V}$,  and the others consist of continuous unions of products $\alpha \times
\beta$
 of periodic orbits $\alpha, \beta$ with
the same (or rationally related) periods (see [U.Z, Lemma 3.7]). Under assumption $Q$, only
`squares' $\alpha \times \alpha$ of periodic orbits arise.
 It follows that each
component contributes with equal strength to the pair correlation function.  

To complete the calculation we need to determine the canonical density on ${\cal P}$ whose
integral gives the principal symbol of $\Upsilon_f$ at $\theta = 0$ (or at other possible
singular angles). On the $T = 0$ component it is just the Liouville density of ${\cal V}$
 and hence the
contribution of this component is the volume of ${\cal V}.$  For the $T > 0$ components we
use the Morse theory of $H$ to break up $M$ into local action-angle charts
$X^{\nu}_j$.   We then wish to determine the surface measures on the surfaces
${\cal V} \cap \{T_j^{\nu}(z_1) = T_j^{\nu}(z_2)\}\subset X_j^{\nu} \times X_j^{\nu}$ where as above $T_j^{\nu}(z)$ is the minimal period of the periodic orbit thru $z$.
We will express the result in terms of the local  action angle coordinates
$(I_1, \phi_1, I_2, \phi_2)$ on $X_j^{\nu} \times X_j^{\nu}$.  The coordinates $(I_1, \phi_1, \phi_2)$ are independent on
${\cal V} \cap \{T_j^{\nu}(z_1) = T_j^{\nu}(z_2)\}$ and (dropping the sub and superscripts) the Liouville measure in these coordinates  equals $T (d I_1 \wedge d\phi_1 \wedge
d\phi_2)$  [U.Z, Lemma 3.11]. On the level sets
$T(z_1) = T(z_2)$ the surface measure is then given by $ T^2 (\frac{d^2 I_1}{d H^2})^{-1} d \phi_1
d\phi_2).$ The measure of the $T > 0$ component of ${\cal P}$ is then the integral over the
set of periods $T$ of this form times $dT.$  Changing variables from $T(E)$ to $E$, summing
over the components $X^{\nu}_j$
(and reinstating the sub and superscripts) gives the
stated formula.
\qed

\section{PCF for quantized perfect Hamiltonian flows on $\C P^1$: Proof of Theorem B}

In this section we consider general Hamiltonians $H$ on $\C P^1$ which are perfect
Morse functions.  Our purpose is to show that the pair correlation functions of
their quantized Hamiltonian flows 
are Poisson on average. 

\subsection{Classical Hamiltonian $S^1$ actions  and perfect Morse Hamiltonians on $\C P^1$}

Up to symplectic equivalence, $\C P^1$ carries a unique Hamiltonian $S^1$ action.  The
model example is that of rotations of $S^2 \subset \Rr^3$ around the $x_3$-axis, whose
 Hamiltonian is the  $x_3$ coordinate.  

In general,  a function $I: \C P^1 \rightarrow \Rr$ is called an `action' variable if
its Hamilton flow $\exp t \Xi_I$  is $2\pi$-periodic.  Any
such global action variable on $\C P^1$ must have precisely two
critical points which are  fixed points of its flow.  This flow
has a global transversal
connecting the fixed points.  The travel time from this transversal defines an
angle variable $\theta$ symplectically dual to $I$ and the transformation $\chi(I,
\theta) = (x_3, \theta)$ is a global symplectic transformation to the model case.

Now  suppose that $H :\C P^1 \rightarrow \Rr$ is a perfect Morse function, with a
non-degenerate minimum value equal to $-1$, and a non-degenerate maximum value
equal to $1$.
Then let $\mu$ denote the distribution function of $H: \C P^1
\rightarrow [-1, 1]$, i.e. $\mu (E) = |\{ H \leq E\}|$ where $|\cdot|$ denotes the
symplectic area. Under our assumptions it is a strictly increasing smooth function
on $(-1,1)$.   Let $I :\C P^1 \rightarrow [0,4\pi]$ be defined by 
$I(z) = \mu (H(z)).$  Then  $I$ is an action variable for $H$: as functions on
the symplectic $\C P^1$,
$I$ and $H$ Poisson commute, $\{I,H\}=0$, and the Hamilton flow of $I$ is $2\pi$-
periodic.  It is obvious that $I$ generates the algebra ${\bf a} = \{f \in
C^{\infty}(\C P^1) : \{f, H\} = 0\}$ and hence we may write $H = \phi (I)$ 
where $\phi$ is the inverse function to $\mu$.
 Non-degeneracy of the critical points forces
 $d\phi \not= 0$ since
\begin{equation} d^2 H = \phi'(I) d^2 I + \phi''(I) dI \otimes dI \end{equation} 
so that $d^2 H = \phi''(I) d I  \otimes dI$ at critical points.

\subsection{Quantum $S^1$ actions over $\C P^1$}

It is a general fact that compact Hamilton group actions can be quantized [B.G].  In the model case,
 the quantization is given by the generator of a linear $S^1$ action, such as rotations around 
the $x_3$-axis (if we think of $\C P^1$ as the embedded $S^2$)
. In general (without conjugation to the model case), 
it can be constructed as follows: The
 Toeplitz quantization $\Pi I \Pi$ of $I$ is a positive operator  satisfying
$e^{2 \pi i {\cal N} [\Pi I \Pi]} = I + K$ where $K$ is a Toeplitz operator of order
$-1$.  In a well-known way [CV], we may add lower order terms to $\Pi I
\Pi$ to arrive at a positive operator $\hat{I}$ satisfying $[\hat{I}, {\cal N}]  =  0$ and
$e^{2 \pi i  {\cal N} \hat{I}} = I.$ Therefore $ {\cal N} \hat{I}$ has only integral eigenvalues.
Since
${\cal N} = N$ on $H^2(N)$, it follows that 
\begin{equation} Sp (\hat{I}|_{H^2(N)}) = \{\frac{j}{N }: j = -N, \dots, N \}.\end{equation}

\subsection{WKB for quantum completely integral Hamiltonians in genus zero}

We now come to the definition of the quantum systems which concern us.  In the homogeneous
setting we say:

\begin{defn} A quantum completely integrable Hamiltonian is a Toeplitz Hamiltonian $\hat{H}$ 
on $H^2(X)$ which is given by a polyhomogeneous function  
$$\hat{H} = \Phi(\hat{I}) \sim \phi(\hat{I})  + \phi_{-1}(\hat{I}) {\cal N}^{-1} +
\phi_{-2}(\hat{I}) {\cal N}^{-2} + \dots$$
 of a global
action operator $\hat{I}$, i.e. a  generator of a quantum Hamiltonian $S^1$ action.  Automatically,
$[\hat{H}, {\cal N}] = 0$ and we write $\hat{H}^{(N)}$ for $\Pi_N \hat{H} \Pi_N.$\end{defn}

The principal symbol $\phi(I)$ defines a perfect Morse function on $\C P^1$ if $\phi' \not= 0.$
Conversely, suppose that $H$ is a perfect Morse function on $M$, 
so that there exists a global action
$I$ and monotone $\phi$ for which $H = \phi(I)$. The Toeplitz quantization $\Pi_N H \Pi_N$ of
$H$ does not necessarily commute with the quantization $\hat{I}$ of $I$, so it is not
quantum completely integrable in the strong sense of the definition above.  To make it commute
it is only necessary to average $\Pi_N H \Pi_N$ with respect to the quantum $S^1$-action
$\exp it \hat{I}.$ Since this does not change the principal symbol it may be regarded as a
quantization of $H$ and we will reserve the notation $\hat{H}$ for it.  Since $\hat{H}$
commutes with $\hat{I}$ and since $\hat{I}$ has a simple spectrum, $\hat{H} = \Phi(\hat{I})$
for some function $\Phi$.  It must be a polyhomogenous function since $\hat{H}$ is a 
Toeplitz operator.  We therefore come back to the same definition as above.
  
The following proposition is an immediate consequence of the polyhomogeneity:

\begin{prop} Denote the eigenvalues of
$\hat{H}^{(N)}$ by  $Sp(\hat{H}^{(N)}) = \{\lambda_{N,j} \}$. Then:
$$\lambda_{N,j} = \phi
 ( \frac{j}{N}) + N^{-1} \phi_{-1}(\frac{j}{N}) +  N^{-2} \phi_{-2}( \frac{j}{N}) + \dots.$$
\end{prop}

\subsection{The quantizations $U_{\chi_t,N}$ and $U_{t,N}$}

There are two approaches to the  quantization of a Hamiltonian flow $\chi_t =\exp t \Xi_H$: (i) by first
exponentiating and then quantizing or (ii) by first quantizing and then exponentiating. As
above we would like to produce a quantum completely integrable system, so that $U_{\chi_t}$
should commute with $\hat{I}$. This could be done by quantizing $H \rightarrow \hat{H}$ as above
and then
exponentiating to get $U_{t, N} = \Pi_N e^{it {\cal N} \hat{H}} \Pi_N.$ This automatically
produces a quantum completely integrable unitary group.

 Alternatively, we could exponentiate $H$ to get the flow
$\exp t \Xi_H$ and then quantize the flow by the Toeplitz method.  By
  Proposition 2.3,  $\exp t \Xi_H$ lifts to a contact transformation
$\chi_t$ on $X$ so  by [Z, \S 3] there exists a canonical  symbol $\sigma_t \in
S^0 (T^*( X))$  such that the  $U_{\chi_t, N} :=
\Pi_N \sigma_t \chi_t \Pi_N$ is unitary.  There is no guarantee that the resulting quantization
commutes with the quantum $S^1$ action generated by $\hat{I}$ nor that it is a unitary group.
The first defect can be overcome, as above, by first averaging against the $S^1$ action and
then applying the functional calculus as in [Z] to make the quantization unitary.
We henceforth use the notation $U_{\chi_t,N}$ to refer to the result.

To compare the methods we prove:

\begin{prop}  $U_{t,N} \cong U_{\chi_t,N}$ modulo operators of order $-1$.
\end{prop}

\noindent{\bf Proof}:  To be cautious, we first convince ourselves  that $\Pi_N$ and $e^{it {\cal N} \Pi H \Pi}$
commute. Indeed, $\Pi_N = \Pi \circ 
\int_{S^1} Ad(e^{i \theta {\cal N}}) e^{-iN \theta} d\theta$ and the averaging
operator commutes with both $\Pi$ and multiplication by $H$.  It follows
that 
$$U_{t,N}^*U_{t,N} = \Pi_N e^{- it {\cal N} \Pi H \Pi} \Pi_N  e^{it {\cal N} \Pi H
\Pi} \Pi_N = \Pi_N e^{- it {\cal N} \Pi H \Pi}  e^{it{\cal N} \Pi H \Pi} \Pi_N =
\Pi_N.$$
Thus, $U_{t,N}$ is unitary on $H^2(N).$

We then observe that $U_t = \Pi e^{it {\cal N} \Pi H \Pi } \Pi = \Pi e^{ it {\cal N} H}
\Pi$ is a unitary group of Fourier-Toeplitz integral operators whose underlying
canonical relation is the lift of the Hamilton flow of $H$ to the symplectic
cone $\Sigma \subset T^*(X) = \{ (x, \xi, r \alpha): r\in \Rr^+\}.$   This follows
very much as in the proof that exponentials of first order pseudodifferential operators
are groups of Fourier integral operators.  Indeed,
$\Pi {\cal N} H \Pi$ is a first order Toeplitz (pseudodifferential) operator
with principal symbol $r H.$  Hence its exponential is a Toeplitz Fourier integral
operator with bicharacteristic flow equal to the Hamilton flow of $r H$ on
$\Sigma$.  By Proposition 2.3, this Hamilton flow is the lift of $\exp t \Xi_H$
to $\Sigma.$  

On the semi-classical (non-homogeneous) level, we note that $U_{t,N} = \Pi_N  e^{it {\cal N} \Pi H \Pi }
\Pi_N$  forms a non-homogeneous 
 Hermite-Fourier integral distribution as $N$ varies.
By non-homogeneous is meant an oscillatory integral (with complex phase) associated to a 
 non-homogeneous Lagrangean; here it equals the graph of $\chi_t$ on
$X$.  Both $U_{t,N}$ and $U_{\chi_t,N}$ are Hermite-Fourier integral distributions associated
to the same Lagrangean, the graph of $\chi_t$, and both have
  principal symbols  equal to the graph
1/2-density.  It follows that they can only differ by operators in the same class 
and of order -1.  In fact both are functions of $\hat{I}$ of the form $\exp (i \Phi_t(\hat{I}).$
The only possible difference is in the subprincipal terms in the polyhomogeneous expansions
of the $\Phi$ functions. \qed
\medskip

It seems most natural to concentrate on the unitary groups $U_{t, N}$ so we will
always assume our quantum maps have the form $  \exp i t
\Phi(\hat{I})$ for some polyhomogeneous $\Phi.$  It is straightforward to extend them
to the more general quantum maps $U_{\chi_t,N}$.

\subsection{Final form of $Tr U_{t, N}^{\ell}.$}

It follows from the above that the pair correlation function for the completely integrable
quantum map $\exp(it {\cal N} \phi(\hat{I}))$ is given by 
$$\rho_{2,t}^{(N)}(f) = \hat{f}(0) + \frac{1}{N^2} \sum_{\ell \not= 0} \hat{f}(\frac{2 \pi \ell}{N})
|S_t(N,\ell)|^2$$
where $\phi' \not= 0$ and where
\begin{equation} S_{t} (N, \ell) =  \sum_{j = 1}^N  e^{i  t \ell N \phi (\frac{j}{N})}.
\end{equation}
We will refer to this as the `homogeneous' case since the Hamiltonian has the form
 $\hat{H}^{(N)} = N \phi(\hat{I}^{(N)}).$

In the  general polyhomogeneous case,  the exponential sum
has the form  
 \begin{equation} S_{t} (N, \ell) =  \sum_{j = 1}^N  e^{i  t \ell N \Phi (\frac{j}{N},N)}
\end{equation}
with $\Phi(\frac{j}{N} ,N) = \phi(\frac{j}{N}) + \frac{\phi_{-1}(\frac{j}{N})}{N} + \dots.$
For the sake of simplicity we often assume the Hamiltonian is homogeneous, but we would like
to note that they also hold  in the general case since the lower order terms do not affect
the lattice point counting arguments that are the heart of the matter.   

As regards the individual sums we also note
that  the terms of order
$\frac{\phi_{-3}}{N^3}$ or lower in
$\Phi(j,N)$ cannot  affect the pair correlation function.  That is, let us
put:
\begin{equation}  Z_t(N,\ell) =  \sum_{j = 1}^N  e^{i  t \ell N \Phi_2 (\frac{j}{N},
N) }\end{equation}  with $\Phi_2 (\frac{j}{N}, N) := [\phi (\frac{j}{N}) + 
\frac{\phi_{-1}(\frac{j}{N})}{N} +
\frac{\phi_{-2}(\frac{j}{N})}{N^2}].$  The following is obvious:

\begin{prop} For $f$ with $\hat{f} \in C_o^{\infty}(\Rr)$, 
$$\frac{1}{N}\sum_{\ell \not= 0} \hat{f}(
\frac{2 \pi \ell}{N}) \frac{1}{N} |\{ |S_{\ell}(N,t)|^2 - |Z_{\ell}(N,t)|^2\}| =
O(1/N).$$\end{prop}

\subsection{Proof of Theorem B: pair correlation on average}

We now come to the main results.  In the first we allow for general polyhomogeneous exponents.

\begin{theo} Let  $U_{t, N} = \exp(it {\cal N} \Phi(\hat{I}))$ as above, with
 $|\phi'(x)| \geq C_1$.  Then the limit pair
correlation function $\rho_{2,t}$ for $U_{t,N}$ is Poisson {\em on average}:
$$\frac{1}{b-a} \int_a^b \rho_{2,t}(f) dt = f(0) + \hat{f}(0)$$ for any interval $[a,b]
\subset \Rr$ and any $f \in C_o^{\infty}.$ \end{theo}

\noindent{\bf Proof}: 
By the above, it  suffices to show that 
$$\lim_{N \rightarrow \infty} \frac{1}{b-a}\int_a^b \frac{1}{N^2}\sum_{\ell \not= 0}
\hat{f}(
\frac{2 \pi \ell}{N})   |S_t(N,\ell)|^2 dt = f(0). $$
To prove this, we use the Hilbert inequality  (cf. [Mo, \S 7.6 (28)]) 
$$\frac{1}{b-a} \int_a^b |\sum_{j=1}^N  e^{2 \pi i \mu_j t}|^2 dt =
N + O(\sum_{j = 1}^N \frac{1}{\delta_j})$$
with $$\delta_j = \min_{1\leq j \leq N, j \not= k} |\mu_j - \mu_k|.$$
The two terms correspond respectively to the diagonal and to the off-diagonal
in the square, and the O-symbol is an absolute constant (which can be taken to
be 3/2).
In the case at hand, $\mu_j = N \ell [ \phi(\frac{j}{N}) + \frac{\phi_{-1}(\frac{j}{N})}{N} 
+ \frac{\phi_{-2}(\frac{j}{N})}{N^2} ] $ so that
$$\delta_{N,j} \geq \ell \;\;\; [ min_{x \in [a,b]} |\phi'(x)| + O(1/N)].$$ 
It follows that if $|\phi'(x)| \geq C > 0$ then  
$$\frac{1}{b-a}\int_a^b    |S_t(N,\ell)|^2  dt = N + O(\frac{N}{\ell}).$$
Therefore the limit equals   
$$\lim_{N \rightarrow \infty}  \frac{1}{N}\sum_{\ell \not= 0} \hat{f}(
\frac{2 \pi \ell}{N}) +  
O(\limsup_{N \rightarrow \infty} \frac{1}{N}\sum_{\ell \not= 0} \frac{1}{\ell}
\hat{f}( \frac{2 \pi \ell}{N})).$$
The first term tends to $\int_{\Rr} \hat{f}(x) dx = f(0).$ When supp $\hat{f} \subset
[-C, C]$ the second
is
$$ << 
\frac{1}{N} [\sum_{\ell \leq C N} \frac{1}{\ell}] << \frac{\log N}{N}.$$\qed
\medskip

\subsection{Proof of Theorem B (a): Number variance on average}

Next, we prove that the number variance is Poisson on average:

\begin{theo} With the same hypotheses as above, we have
$$\lim_{N \rightarrow \infty} \frac{1}{b-a}\int_a^b \Sigma_{2,t}^{(N)} dt = L.$$  
\end{theo}

\noindent{\bf Proof}

We have 
\begin{equation} \frac{1}{b-a}\int_a^b \Sigma_{2,t}^{(N)} dt =  \frac{2}{\pi^2}
\sum_{\ell = 1}^{\infty} \frac{1}{\ell^2} \sin^2(\frac{\pi \ell L}{N})
\frac{1}{b-a}\int_a^b  |Tr U_{t, N}^{\ell}|^2 dt  \end{equation}
 
\begin{equation} =  N \frac{2}{\pi^2}
\sum_{\ell = 1}^{\infty} \frac{1}{\ell^2} \sin^2(\frac{\pi \ell L}{N}) + 
\frac{2}{\pi^2}
\sum_{\ell = 1}^{\infty} \frac{1}{\ell^2} \sin^2(\frac{\pi \ell L}{N})\;\;\{
\frac{1}{b-a}\int_a^b \sum_{j \not = k, j,k=1}^N e^{2\pi i N \ell
(\phi(\frac{j}{N}) - \phi(\frac{k}{N}))}  dt\}. \end{equation}
The first term is a Riemann sum and we have 
$$\frac{1}{N} \frac{2}{\pi^2}
\sum_{\ell = 1}^{\infty} \frac{N^2}{\ell^2} \sin^2(\frac{\pi \ell L}{N}) \rightarrow
\int_0^{\infty} (\frac{\sin L x}{x})^2 dx = L.$$
The second can be explicitly evaluated as above and is bounded by
$$N \log N \sum_{\ell = 1}^{\infty} \frac{1}{\ell^3} \sin^2(\frac{\pi \ell L}{N}) =
O(\frac{(\log N)^2}{N}).$$ \qed

\section{Mean square Poisson statistics for quantum spin evolutions: Proof of 
Theorem B (b) }

We have just seen that averages of the pair correlation function
$\rho_{2,t}^{(N)}$ of quantized Hamilton flows $U_{t,N}$ converge to
$\rho_2^{POISSON}$ as long as $H$ is a perfect Morse function.  In particular,
the result is true for linear functions $H = \alpha I$ in the action.  Since
exponential sums with linear phases are far from random (see \S 6), it is evident
that the averaging is the agent producing the random number behaviour.  

A much stronger test of the Poisson behaviour of $U_{t,N}$ is whether the variance
$$\frac{1}{b - a} \int_a^b |\rho_{2,t}^{(N)}(f) - \rho_2^{POISSON}(f)|^2 dt $$
tends to zero as $N \rightarrow \infty$.  It is easy to see that quantum Hamiltonian
flows with linear Hamiltonians do not have this property (\S 4.1).  We therefore
turn  to  quadratic Hamiltonians $\hat{H} = \alpha \hat{I}^2 + \beta \hat{I}.$  Our next result shows
that the pair correlation functions of their quantized Hamilton flows converge  in mean square 
 to Poisson. The proof is based on techniques from mean value estimates on exponential sums; see
[B.I] [G.K] [H.1] for background.

\begin{theo} Let $\hat{H}_{\alpha, \beta} = \alpha \hat{I}^2 + \beta \hat{I}$ and let
 $\rho_{2;(t, \alpha, \beta)}^{(N)}$ be the pair correlation measure for
the quantum map $U_{(t,\alpha, \beta), N} = exp(i t {\cal N} \hat{H}_{(\alpha,
\beta; N)} ).$  Then for any $t \not= 0$, any $T > 0$ and any $f \in {\cal S}(\Rr)$
with $\hat{f} \in C_o^{\infty}(\Rr)$ we
have $$ \frac{1}{(2T)^2} \int_{-T}^T \int_{-T}^T
|\rho_{2;(t, \alpha, \beta)}^{(N)}(f) - \rho_{2}^{POISSON}(f)|^2 d\alpha d\beta = 0
(\frac{(\log N)^2}{N}).$$ 
\end{theo}

\noindent{\bf Proof}: Removing the $\ell = 0$ and diagonal terms as above, it suffices
to show that 
\begin{equation}\frac{1}{(2T)^2} \int_{-T}^T
\int_{-T}^T |\frac{1}{N^2} \sum_{\ell \not= 0} \hat{f}(\frac{\ell}{N}) [\sum_{j \not=
k, j,k = 1}^N e(t (\alpha \frac{j^2 - k^2}{N} + \beta(j - k))) |^2 d\alpha d\beta =
0(\frac{(\log N)^2}{N}).\end{equation}
To prove this, we use the Beurling - Selberg function $B_{T, |t|\delta}.$
It has  the following properties (see [G.K]):

\noindent$\bullet$ $B_{T, |t| \delta} \geq \chi_{[-T, T]}$;\\
$\bullet$ Supp $\hat{B}_{T, |t| \delta} \subset (- |t| \delta, |t| \delta).$

Here, $\chi_{[-T, T]}$ is the characteristic function of $[-T, T].$  Then  the
integral on the right side above is bounded above by 
\begin{equation} \int_{\Rr}
\int_{\Rr} B_{T,|t|\delta}(\alpha) B_{T, |t|\delta} (\beta) |\frac{1}{N^2} \sum_{\ell
\not= 0} \hat{f}(\frac{\ell}{N}) [\sum_{j \not= k, j,k = 1}^N e(t (\alpha \frac{j^2 -
k^2}{N} + \beta(j - k))) |^2 d\alpha d\beta. \end{equation}
Squaring and evaluating the Fourier transforms gives
  \begin{equation}  \frac{1}{N^4} \sum_{\ell_1 \not= 0} \sum_{\ell_2 \not= 0}
\hat{f}(\frac{\ell_1}{N}) 
\bar{\hat{f}}(\frac{\ell_2}{N})  [\sum_{j_1 \not= k_1, j_1,k_1 = 1}^N  
\sum_{j_2 \not= k_2, j_2,k_2 = 1}^N  \hat{B}_{T,
|t|\delta}(t (\ell_1 \frac{j_1^2 - k_1^2}{N} - \ell_2 \frac{j_2^2 - k_2^2}{N}))
\hat{B}_{T, |t| \delta} (t (\ell_1 (j_1 - k_1) - \ell_2(j_2 - k_2))) . \end{equation}

By the support properties of $B_{T,  |t| \delta}$, the latter expression is bounded
above by 
  \begin{equation}  \frac{1}{N^4} \sum_{\ell_1 \not= 0} \sum_{\ell_2 \not= 0}
|\hat{f}(\frac{\ell_1}{N})| 
|{\hat{f}}(\frac{\ell_2}{N})| I(N, \ell_1, \ell_2) \end{equation}
with
\begin{equation} I(N, \ell_1, \ell_2) = \# \{(j_1, k_1, j_2, k_2) \in [1, N]^4: j_i
\not= k_i : 
| (\ell_1 \frac{j_1^2 - k_1^2}{N} - \ell_2 \frac{j_2^2 - k_2^2}{N})| \leq \delta,
| (\ell_1 (j_1 - k_1) - \ell_2(j_2 - k_2)))| \leq \delta \} . \end{equation}
Introduce new variables $h_i = j_i - k_i, m_i = j_i + k_i$ so that the conditions
read
$$\left(\begin{array}{l} |\ell_1 \frac{h_1 m_1}{N} - \ell_2 \frac{h_2 m_2}{N})|
\leq \delta \\
\\
| \ell_1 h_1 - \ell_2  h_2| \leq \delta \end{array}\right).$$
The change of variables is invertible so $I(N, \ell_1, \ell_2)$ is the number
of integer solutions $(h_1, h_2, m_1, m_2)$ with $|m_i| \leq 2N, |h_i| \leq N -
|m_i|.$  

Since $| \ell_1 h_1 - \ell_2  h_2| \in {\bf N}$ it can only be $< \delta$ if it
 vanishes.  Therefore the second condition is equivalent to
$$\langle \ell, h \rangle = 0 \;\;\;\Rightarrow h_2 = \frac{\ell_1}{\ell_2} h_1.$$
Here we assume $|h_1|\geq |h_2|$ so that $|h_1| = max \{|h_1|, |h_2|\} \sim |h|$
and we abbreviate $\ell = (\ell_1, \ell_2)$ etc.  Substituting in the first condition
we get
$$\ell_1 h_1 (m_1 - m_2) = O(N \delta).$$

 Now let us count solutions. We split them up into two classes: (a) homogeneous solutions
with $m_1 - m_2 = 0$ and (b) inhomogeneous solutions with $m_1 - m_2 \not= 0.$  The
homogeneous solutions are lattice points $(h_1, m_1, \ell_1; h_2, m_2, \ell_2)$ which
solve the system
$$\left\{ \begin{array}{l} \ell_1 h_1 = \ell_2 h_2 \;\;\;\;\;(h_1 \not= 0, h_2 \not= 0); \\ \\
m_1 = m_2. \end{array} \right.$$
Clearly there are $2N$ solutions of $m_1 = m_2$. For each integer $s$ in $[-N^2, N^2]$
there are $ \leq d(s)$ ways of writing $s$ as a product $h_1 \ell_1$ with $h_1, \ell_1 \in
[-N, N].$ Here, $d(s)$ is the divisor function (the number of non-trivial divisors of
$s$). Hence the number of solutions of the homogeneous system is $O(N \sum_{s=1}^{N^2}
d(s)^2) = O(N^3 (\log N)^3).$

Now let us count inhomogeneous solutions.  
We write $I_{ih}(N, \ell_1, \ell_2)$ for the number of inhomogeneous solutions
with fixed $(\ell_1, \ell_2).$  
 Since $h_2$ is determined from $(h_1, m_1, m_2)$
it suffices to count these triples. 
  First,
there are $O(N)$ choices of $m_1.$  Then put $m_2 = m_1 + M$ with $M \geq 1$ so that $h_1 M =
O(\frac{N}{\ell_1} \delta)$.   From $M \leq 2N$ the number of pairs $(h_1, M)$ is
bounded above by $$\sum_{M = 1}^{2 N} \frac{N}{\ell_1} \frac{1}{M} =
O(\frac{N}{\ell_1} \log N).$$ 
Hence $I_{ih}(N, \ell_1, \ell_2) <<  \frac{N^2 \log N}{\ell_1 }.$

It follows that for $\hat{f} \in C_o^{\infty}(\Rr)$, 
  \begin{equation}\begin{array}{l}  \frac{1}{N^4} \sum_{\ell_1 \not= 0} \sum_{\ell_2 \not= 0}
|\hat{f}(\frac{\ell_1}{N})| 
|\hat{f}(\frac{\ell_2}{N})| I(N, \ell_1, \ell_2) << O(\frac{(\log N)^3)}{N}) + 
 \frac{\log N}{N^2 } \sum_{\ell_1 \not= 0} \sum_{\ell_2 \not= 0}
|\hat{f}(\frac{\ell_1}{N})| 
|\hat{f}(\frac{\ell_2}{N})| \frac{1}{\ell_1 } \\ \\
<< O(\frac{(\log N)^3)}{N})+ O(\frac{(\log N)^2}{N}) = O(\frac{(\log N)^3)}{N}). \end{array}
 \end{equation}

\qed

\subsection{Proof of Theorem B(b) for general non-degenerate phases}

We now consider a much more general class of Hamiltonians for which a similar result holds:

\begin{theo} Let $\hat{H}_{\alpha, \beta} = \alpha \phi(\hat{I}) + \beta \hat{I}$ where $|\phi''| > C > 0$
on $[-1, 1]$ and let
 $\rho_{2;(t, \alpha, \beta)}^{(N)}$ be the pair correlation measure for
the quantum map $U_{(t,\alpha, \beta), N} = exp(i t {\cal N} \hat{H}_{(\alpha,
\beta; N)} ).$  Then for any $t \not= 0$, any $T > 0$ and any $f \in {\cal S}(\Rr)$
with $\hat{f} \in C_o^{\infty}(\Rr)$ we
have $$ \frac{1}{(2T)^2} \int_{-T}^T \int_{-T}^T
|\rho_{2;(t, \alpha, \beta)}^{(N)}(f) - \rho_{2}^{POISSON}(f)|^2 d\alpha d\beta =
 0(\frac{(\log N)^2}{N}).$$ 
\end{theo}

\noindent{\bf Proof}~~~ The previous argument now leads  to the lattice point
problem:
\medskip

$$\left( \begin{array}{l} \ell_1 h_1 = \ell_2 h_2 \\
\\
\ell_1 (\phi(\frac{j_1}{N}) - \phi(\frac{k_1}{N})) - \ell_2 (\phi(\frac{j_2}{N}) -
 \phi(\frac{k_2}{N})) = O(1/N) \end{array} \right)$$ 
\medskip

Here as above $h_i = j_i - k_i$.
By the mean value theorem there exist $\xi_{j_i k_i} \in [j_1, k_1]$ such that
$\phi(\frac{j_i}{N}) - \phi(\frac{k_i}{N}) =\frac{1}{N} \phi'(\xi_{j_i k_i}/N) (j_i - k_i).$
As above we then get the system of constraints:
\medskip

$$\left( \begin{array}{l} \ell_1 h_1 = \ell_2 h_2 \\
\\
h_1 \ell_1 (\phi'(\xi_{j_1 k_1}/N) - \phi'(\xi_{j_2 k_2}/N)) = O(1) \end{array} \right) $$
\medskip

Then writing $\phi'(\xi_{j_1 k_1}/N) - \phi'(\xi_{j_2 k_2}/N) = \phi''(\xi_{j_1 k_1 j_2 k_2}/N)
(\xi_{j_1 k_1} - \xi_{j_2 k_2})/N$ we get the system:
\medskip

$$\left( \begin{array}{l} \ell_1 h_1 = \ell_2 h_2 \\
\\
h_1 \ell_1 \phi''(\xi_{j_1 k_1 j_2 k_2}/N) (\xi_{j_1 k_1} - \xi_{j_2 k_2}))= O(N) \end{array}
\right)$$
\medskip

By the assumption $|\phi''| > c > 0$  this gives 
 
\begin{equation}\label{SYSTEM} \left( \begin{array}{l} \ell_1 h_1 = \ell_2 h_2 \;\;\;\;\;\;(\ell_j \not= 0, h_j \not= 0)\\
\\
h_1 \ell_1  (\xi_{j_1 k_1} - \xi_{j_2 k_2})= O(N) \end{array} \right. \end{equation}
Let us change to the variables  $(h_i, m_i)$ as in the quadratic case  and write
$\xi_{j_i k_i} = \xi(h_i, m_i)$. 
 We wish to count the number of lattice points $(\ell_1, h_1, m_1, \ell_2, h_2, m_2)$
satisfying the system 
\begin{equation}\label{NEWSYSTEM} \left( \begin{array}{l} \ell_1 h_1 = \ell_2 h_2 \;\;\;\;\;\;(\ell_j \not= 0, h_j \not= 0)\\
\\
  |\xi(h_1,  m_1) - \xi(h_2,  m_2)|\leq \frac{N}{h_1 \ell_1}. \end{array} \right.\end{equation}

 As in the quadratic case, we regard $(h_1, \ell_1, m_1)$ as independent variables,
so that the first equation is a constraint on $(h_2, \ell_2).$ 
We also regard the second constraint $  |\xi(h_1,  m_1) - \xi(h_2,  m_2)|\leq \frac{N}{h_1 \ell_1}$ as a constraint on $|m_2|.$   To put it in a more convenient form we consider
$\xi(h_2, m_2)$ as a function $\xi_{h_2}(m_2)$ and   invert the
function $\xi_{h_2}.$  An easy calculation gives
\begin{equation} \xi_{h} (m) = 
N (\phi')^{-1} (\int_0^1 \phi'( \frac{m}{2N} + (2s-1) \frac{h}{2N}) ds )\end{equation}
hence
\begin{equation} \frac{\partial}{\partial m} \xi_{h}(m) =
 [(\phi')^{-1}]'\{ (\int_0^1 \phi'( \frac{m}{2N} + (2s-1) \frac{h}{2N}) ds )\}
[\int_0^1 \phi''( \frac{m}{2N} + (2s-1) \frac{h}{2N}) ds].\end{equation}
Since $C \leq |\phi''(x)| \leq C'$ for $x \in [-1,1]$ and certain positive
constants $C, C'$,  It follows that 
that  $|\frac{\partial}{\partial m} \xi_{h}(m)| \geq \delta > 0 \;\;\;\;(\forall
m \in [0, 2N]$).  Therefore a smooth inverse function $\xi_h^{-1}$ exists on the range
of $\xi$.   In particular, $\xi_h^{-1}$ is Lipshitz, so (\ref{NEWSYSTEM}) is equivalent
to the system
\begin{equation}\label{NEWERSYSTEM} \left( \begin{array}{l} \ell_1 h_1 = \ell_2 h_2 \;\;\;\;\;\;(\ell_j \not= 0, h_j \not= 0)\\
\\
|m_2 - \xi_{h_2}^{-1}(\xi(h_1,  m_1))| \leq C' \frac{N}{h_1 \ell_1}. \end{array} \right)\end{equation}
The situation is now very close to the quadratic case: There are $d(h_1 \ell_1)$
solutions $(h_2, \ell_2)$ of the first equation and $N$ values of
$m_1$.  For given $(h_1, \ell_1, h_2, \ell_2, m_1)$ there are $\leq 1 +
 O( \frac{N}{h_1 \ell_1})$ solutions $m_2$  of the second equation.
Summing $1$ over the relevant  $(h_1, \ell_1, h_2, \ell_2, m_1)$
 gives $O(N^3 (\log N)^3$ as in the homogeneous part of the
quadratic case.  Summing $O( \frac{N}{h_1 \ell_1}$ over these lattice points gives 
$O(N^2 (\log N)^2)$ as in the inhomogeneous part of the
quadratic case.  
 \qed

\begin{cor} Let $N_m = [m (\log\; m)^4] $ ($[ \cdot ]$ = integer part). Then for almost all
$(\alpha, \beta)$ with respect to Lebseque measure and all $t \not= 0$ we have
$$\lim_{m \rightarrow \infty} \rho_{2; (t, \alpha, \beta)}^{N_m} = \rho_2^{POISSON}.$$ 
\end{cor} 

\noindent{\bf Proof} By the above,
$$\sum_{m=1}^{\infty}  \frac{1}{(2T)^2} \int_{-T}^T
 \int_{-T}^T 
|\rho_{2;(t, \alpha, \beta)}^{(N_m)}(f) - \rho_{2}^{POISSON}(f)|^2 d\alpha d\beta < \infty.$$
Since the terms are positive it follows that for almost all $(\alpha, \beta)$, 
$$\sum_{m=1}^{\infty} |\rho_{2;(t, \alpha, \beta)}^{(N_m)}(f) - 
\rho_{2}^{POISSON}(f)|^2 d\alpha d\beta < \infty$$
and for these $(\alpha, \beta)$ the $m$th term tends to zero. The set of such $(\alpha, \beta)$
apriori depends on the smooth function $f$. However by a standard diagonal argument one can find
a set of full measure that works for every continuous $f$ (see [Sa.2][R.S] for further details).
\medskip

\noindent{\bf Remark}  In this corollary we have adapted an argument from
  [Sa.2][R.S], where the pair correlation problem is studied for flat
tori and for some homogeneous integrable systems.  Their main result was that the relevant
 pair correlation functions are almost everywhere Poisson.  After proving the almost everywhere
convergence to Poisson along a slightly sparse subsequence (as in the above Corollary), they show that
for 
$N_m < M < N_{m+1}$, $$\rho_{2;(t, \alpha, \beta)}^{(N_m)}(f) - \rho_{2;(t, \alpha, \beta)}^{(M)}(f)
= o(1)$$ as $m \rightarrow \infty$.  This last step seems to be much more difficult in our 
problem. The difference is that the spectra in  [Sa.2][R.S]  increase with increasing $N$ and the
common terms cancel in the difference above.  On the other hand, our spectra change rapidly with
$N$ and there are no (obvious) common terms to cancel.  In [Z.Addendum] we will show that an
a.e. result comparable to that of [R.S] holds for the average in $N$ of
 $\rho_{2;(t, \alpha, \beta)}^{(N)}$ if the phase is a polynomial.

\section{Appendix: Linear and quadratic cases}

In the case of linear and pure quadratic Hamiltonians, the exponential sums discussed above
are classical and there are many prior results in the literature.
  We briefly discuss what is known and add a few observations of our own.

First, the pair correlation problem for linear Hamiltonians $H = \alpha I$ has been studied
since the fifties.  See [Bl.2][R.S][Ca.Gu.Iz] for discussion and
references to the literature. The main result is that only three level spacings can occur
 for a given $\alpha$ and  the
pair correlation function is not even mean square Poisson.  

In the case of quadratic Hamiltonians, we get the incomplete Gauss sums:
$$S_t(N; \ell) = \sum_{j=1}^N e^{2\pi i t N \ell [(\frac{j}{N})^2 + 
\alpha \frac{j}{N}] }.$$
In the special case
 $\alpha = 0$ and $t = 1$ they are classical complete Gauss sums
 $$G(\ell, 0; N) \sum_{j=1}^N e^{2\pi i  \ell  \frac{j^2}{N}}.$$
If $ (\ell , N) = 1$ then
$$|G(\ell, 0, N)| = \left\{ \begin{array}{ll} \sqrt{N} & if N \equiv 1 (mod 2) \\
\sqrt{2N} & if N \equiv 0 (mod 4)\\
0 & if N \equiv 2 (mod 4) 
\end{array} \right .$$
In general
$$G(\ell, 0, N) = (\ell, N) G(\frac{\ell}{(\ell, N)}, 0; \frac{N}{(\ell, N)}).$$
Hence the values of 
$$I_{N} = \frac{1}{N^2}\sum_{\ell \not= 0}
\hat{f}(
\frac{2 \pi \ell}{N})   |S_t(N,\ell)|^2$$
depend on the residue class of $N$ modulo 4.  If $N \equiv 2$ (mod 4) then 
$I_{N} = 0.$ If $N$ is odd, then   
$$I_{N} = \frac{1}{N}\sum_{\ell \not= 0} (\ell, N)
\hat{f}( \frac{2 \pi \ell}{N}) = 
\frac{1}{N} \sum_{k \in \Z: k \not= 0} k \sum_{\ell: (\ell, N) = k}
\hat{f}(\frac{2 \pi \ell}{N}).$$

When $N = p$, a prime number. Then $(\ell, p) = 1$ except for
multiples $k p$ with $k \in supp(\hat{f}).$ They make a neglible contribution, so
$$I_{p} = \frac{1}{p}  \sum_{\ell \not= 0}
\hat{f}(\frac{2 \pi \ell}{p})\rightarrow \int_{\Rr} \hat{f}(x) dx = f(0).$$
Thus the prime sequence is Poisson.

At the opposite extreme, suppose $N = p^M$ for some prime $p$. Then
$$I_{p^M} = \frac{1}{p^M} 
\sum_{k = 0}^M p^k \sum_{q = 1, (q,p) = 1}^{p^{M - k}} \hat{f}(\frac{q}{p^{M-k}}).$$
The sums
$$ \frac{1}{p^{M - k}} \sum_{q = 1, (q,p) = 1}^{p^{M - k}} \hat{f}(\frac{q}{p^{M-k}})$$
have the form of Riemann sums for $f(0)$ as $M - k \rightarrow \infty$ except that the
terms with $p | q$ are omitted.  These terms also resemble Riemann sums for $f(0)$, multiplied
by $1/p$.  Hence each term is roughly $1 - 1/p$ times $f(0).$  Since there are $M$ such terms,
the coefficient of $f(0)$ tends to infinity and the pair correlation function cannot be
Poisson.

\subsection{Poisson on average in $t$}

If we allow $t$ to vary then we do have an average Poisson behaviour:

\begin{prop} For any interval $[-T,T]$, the average PCF of $exp it {\cal N} \hat{I}^2$
is Poisson, i.e.
$$\frac{1}{N^2} \sum_{\ell \not= 0} \hat{f}(\frac{\ell}{N}) \frac{1}{2T}
\int_{-T}^T \sum_{j \not= k} e^{i \ell t \frac{j^2 - k^2}{N}} dt = o(1).$$\end{prop}
\medskip

\noindent{\bf Proof}: The integral equals  
$$\frac{1}{N} \sum_{\ell \not= 0} \frac{1}{\ell} \hat{f}(\frac{\ell}{N})
 \sum_{j \not= k} \frac{sin ( \ell T \frac{j^2 - k^2}{N})}{j^2 - k^2}$$
$$= \frac{1}{N^2} \sum_{\ell \not= 0} \hat{f}(\frac{\ell}{N})
[ \sum_{m=1}^{2N} \sum_{0 <|h| \leq N - |m|} \frac{\sin ( \ell T \frac{hm}{N})}
{N h m \ell }$$
where as above $h = j-k, m= j+k.$ Using just that $\sin x << 1$ this is
$$<<\frac{1}{N} \sum_{\ell \not= 0} \hat{f}(\frac{\ell}{N})
 \sum_{m=1}^{2N} \sum_{0 <|h| \leq N - |m|} \frac{1} 
{ h m \ell }$$
$$<<\frac{(\log N)^2}{N} \sum_{\ell \not= 0} \hat{f}(\frac{\ell}{N}) 
\frac{1}{\ell} = O(\frac{(\log N)^2}{N}).$$
\qed
\medskip

We have not determined whether  the variance tends to zero in this case.

\end{document}